\documentclass{article}

\usepackage[preprint]{neurips_2024}

\usepackage[utf8]{inputenc} % allow utf-8 input
\usepackage[T1]{fontenc}    % use 8-bit T1 fonts
\usepackage[hidelinks]{hyperref}       % hyperlinks
\usepackage{url}            % simple URL typesetting
\usepackage{booktabs}       % professional-quality tables
\usepackage{amsfonts}       % blackboard math symbols
\usepackage{nicefrac}       % compact symbols for 1/2, etc.
\usepackage{microtype}      % microtypography
\usepackage{xcolor}         % colors

\usepackage{orcidlink}
\usepackage{bm}
\usepackage{mathtools}
\usepackage{amsmath}
\usepackage{wrapfig}
\usepackage{tikz}
\usepackage{multirow}
\usepackage{color}
\usepackage{xcolor}
\usepackage{url}
\usepackage{caption}
\usepackage{subcaption}
\usepackage{float}
\usepackage{balance}
\usepackage{graphicx}
\usepackage{textcomp}
\usepackage{booktabs}
\usepackage{colortbl}
\usepackage{enumerate}
\usepackage{algorithm}
\usepackage[noend]{algpseudocode}

\def\eps{{\epsilon}}
\usepackage{bbm}

\newcommand\shortsection[1]{\vspace{6pt}{\noindent\bf #1.}}

\DeclareMathOperator*{\argmin}{arg\,min}
\def\E{\mathop{\mathbb{E}}}

\newcommand{\EE}{\mathbb E}

\title{Transferable Availability Poisoning Attacks}

\author{
\vspace{1mm}
Yiyong Liu \quad Michael Backes \quad Xiao Zhang \\
\vspace{1mm}
CISPA Helmholtz Center for Information Security \\
\texttt{\{yiyong.liu,director,xiao.zhang\}@cispa.de}
}

\begin{document}

\maketitle

\begin{abstract}
We consider availability data poisoning attacks, where an adversary aims to degrade the overall test accuracy of a machine learning model by crafting small perturbations to its training data. Existing poisoning strategies can achieve the attack goal but assume the victim to employ the same learning method as what the adversary uses to mount the attack. In this paper, we argue that this assumption is strong, since the victim may choose any learning algorithm to train the model as long as it can achieve some targeted performance on clean data. Empirically, we observe a large decrease in the effectiveness of prior poisoning attacks if the victim employs an alternative learning algorithm. To enhance the attack transferability, we propose \textit{Transferable Poisoning}, which first leverages the intrinsic characteristics of alignment and uniformity to enable better unlearnability within contrastive learning, and then iteratively utilizes the gradient information from supervised and unsupervised contrastive learning paradigms to generate the poisoning perturbations. Through extensive experiments on image benchmarks, we show that our transferable poisoning attack can produce poisoned samples with significantly improved transferability, not only applicable to the two learners used to devise the attack but also to learning algorithms and even paradigms beyond.
\end{abstract}

\section{Introduction}
\label{sec:intro}

With the growing need for utilizing large amounts of data to train machine learning models, especially for training state-of-the-art large-scale models, online data scraping has become a widely used tool. However, the scraped data often come from untrusted third parties, which undesirably empowers adversaries to execute data poisoning attacks more easily. Availability poisoning~\cite{HMEBW21,FGCGCG21,FHLST22,YZCYL22,SSGGGJ22,HZK23,ZMYSJWX23}, a specific type of data poisoning attack, has received a lot of attention recently, due to their potential threats to the robustness of model training pipelines. More specifically, an availability poisoning attack aims to reduce the model test performance as much as possible by injecting carefully crafted imperceptible perturbations into the training data. Existing attack strategies are successful in generating poisoned data samples that are highly effective in lowering the model test accuracy with respect to a given supervised~\cite{HMEBW21,FGCGCG21} or unsupervised learner~\cite{HZK23,ZMYSJWX23}.
However, the effectiveness of these attacks largely relies on the assumption that the victim employs the same learning method to train the model as the reference learner that the adversary uses to devise the attack. 

We argue that imposing such an assumption is unreasonable, as the victim has the flexibility to choose from a wide range of learning algorithms to achieve their objectives, especially witnessing the rapid advancement of unsupervised~\cite{CKNH20,CFGH20,GSATRBDPGAPKMV20,CH21} learning methods. In particular, these approaches can often match or even surpass the performance of supervised learning in various machine learning tasks~\cite{CKSNH20,CXH21,RKHRGASAMCKS212}, which greatly expands the range of options for the victim to train a satisfactory model. The advancement of these alternative methods has prompted us to study the transferability of availability poisoning attacks across different learning methods and paradigms. Unfortunately, we observe a significant decrease in the effectiveness of existing attacks when the victim uses a different learning paradigm to train the model, which renders them less effective in relevant real-world applications (see Figure~\ref{figure:motivation} for a heatmap visualizing the transferability of clean model accuracy and the effectiveness of existing attacks across different victim learners).

More specifically, we hypothesize that the challenge in transferring poisoning perturbations primarily stems from the differences in attack mechanisms across various learning algorithms. As demonstrated by previous works, for most attacks in supervised learning~\cite{HMEBW21,FHLST22,YZCYL22,SSGGGJ22}, the generated poisoning perturbations exhibit linear separability, providing a shortcut for the model to converge quickly. However, in unsupervised contrastive learning, the poisoning methods work by aligning two augmented views of a single image and are much less linearly separable~\cite{HZK23,RXWMST23}.
By leveraging the insights gained by investigating the transferability of prior availability attacks across various victim learners, we propose a novel method that can generate poisoning samples with significantly improved transferability across different prominent learning methods.

\shortsection{Contributions}
Our general idea is to search for shared characteristics among different learning algorithms to enhance the transferability of generated poisons. We demonstrate the disparate transferability performance of existing availability attacks with respect to different victim learners, spanning across supervised and contrastive learning (Figure~\ref{figure:motivation}). Among them, contrastive learning is more robust to poisoning attacks, even the poisons generated based on contrastive learning will be overfitted to the same training framework, while with poor transferability to other learning algorithms. As contrastive loss learns similar representations for positive pairs and pushes negative pairs apart, the learned representations should align different views of the same image and preserve as much information as possible, denoted as alignment and uniformity~\cite{WI20} respectively. We first observe that contrastive poisoning with alignment and uniformity can largely improve the transferability within contrastive learning (Figure~\ref{figure:motivation}). Furthermore, to align the poisoning effect across supervised and contrastive learning, we propose to leverage the information from a shared model trained using supervised and contrastive learning with alignment and uniformity iteratively to generate perturbation with an error-minimization scheme. Such a training strategy can be regarded as an approximated way for the poisons to achieve the worst-case unlearnability (Section~\ref{sec:transfer}).

Based on the two steps mentioned above, we propose \textit{Transferable Poisoning} (TP) to generate poisoning perturbations that can be transferred across different supervised and contrastive learning algorithms (Section \ref{sec:transfer}). In order to validate the effectiveness and transferability of our proposed TP, we conduct extensive experiments on image benchmarks under different settings. Our results demonstrate that TP is not only highly effective in the two chosen learning paradigms but also exhibits significantly improved transferability to other related training algorithms (Sections \ref{sec:main_results} - \ref{sec:model_architecture}). For instance, the worst-case unlearnability of the poisoned data produced by our method is as low as 41.72\%, 4.57\% and 9.34\% on CIFAR-10, TinyImageNet and MiniImageNet respectively, exhibiting an improvement of 6.89\%, 23.38\% and 23.69\% in attack effectiveness compared with existing methods (Table~\ref{table:main_results} and~\ref{table:main_results_tiny_mini}). In addition, we test alternative attack methods and provide deeper insights to explain the advantages of our TP (Section~\ref{sec:discussion}). Our attack exhibits superior transferability compared to baselines, demonstrating more threat of availability poisoning in real-world scenarios.

\section{Related Work}
\label{sec:related}
\shortsection{Availability Poisoning}
Availability poisoning attacks, also known as indiscriminate data poisoning attacks, aim to undermine the model's overall performance by maliciously manipulating its training data. A line of existing works consider adversaries who attempt to achieve this goal by injecting a small amount of poisoned samples into the clean training dataset~\cite{KL17,SMSET21,LKY23,SZTE23}. However, the poisoned data of these methods are relatively easier to be distinguished from the normal inputs as the allowed adversarial modifications are usually not restricted to be imperceptible, and they also fail to achieve satisfactory performance when applied to non-convex learning methods such as deep neural networks. Another line of research studies availability poisoning attacks that add imperceptible perturbations to the training data to manipulate the victim learner~\cite{HMEBW21,FGCGCG21,YZCYL22,HZK23,ZMYSJWX23}. In particular, Huang et al.~\cite{HMEBW21} propose to craft unlearnable examples based on a reference model with error-minimization perturbations (EM), while Fowl et al.~\cite{FGCGCG21} propose \textit{Targeted Adversarial Poisoning} (TAP), which generates poisoning perturbations in a reverse way using error-maximization objectives. Several recent works consider leveraging the property of linear separability to create poisoning perturbations more efficiently~\cite{YZCYL22,SSGGGJ22}. In addition to supervised learning, the effectiveness of availability poisoning attacks has also been validated in unsupervised learning. For instance, Zhang et al.~\cite{ZMYSJWX23} propose \textit{Unlearnable Clusters} to address the challenge of agnostic labels, whereas He et al.~\cite{HZK23} focus on studying poisoning attacks for unsupervised contrastive learning methods and propose \textit{Contrastive Poisoning} (CP). 

\shortsection{Transferable Poisoning}
Different from previous works~\cite{ZHLTSG19,AMWKV21} which study the transferability of targeted data poisoning attacks where the goal is to misclassify a targeted test image to a specific class, some recent studies explore the transferability of availability poisoning attacks from various perspectives. Huang et al.~\cite{HMEBW21} and Fowl et al.~\cite{FGCGCG21} show that the poisons generated based on one model architecture can be transferred to another. Similarly, He et al.~\cite{HZK23} consider the scenario where the adversary and victim use different frameworks for unsupervised contrastive learning. \textit{Transferable Unlearnable Examples} (TUE)~\cite{RXWMST23} and \textit{Augmentation-based availability attacks} (AUE and AAP)~\cite{WZG24} are the closest works to ours, as they also consider the transferability in a more generalized manner, encompassing different learning paradigms. Specifically, the former method incorporates linear separability into contrastive unlearnable examples, while the latter starts from supervised poisoning and enhances the data augmentations. Different from these works, our method leverages the intrinsic property of alignment and uniformity in contrastive learning and iteratively utilizes the gradient information from supervised and contrastive learning, thereby generating perturbations that exhibit enhanced attack transferability.

\section{Transferable Poisoning}
\label{sec:transfer}

\begin{wrapfigure}{R}{0.4\textwidth}
\vspace{-12pt}
\includegraphics[width=0.4\textwidth]{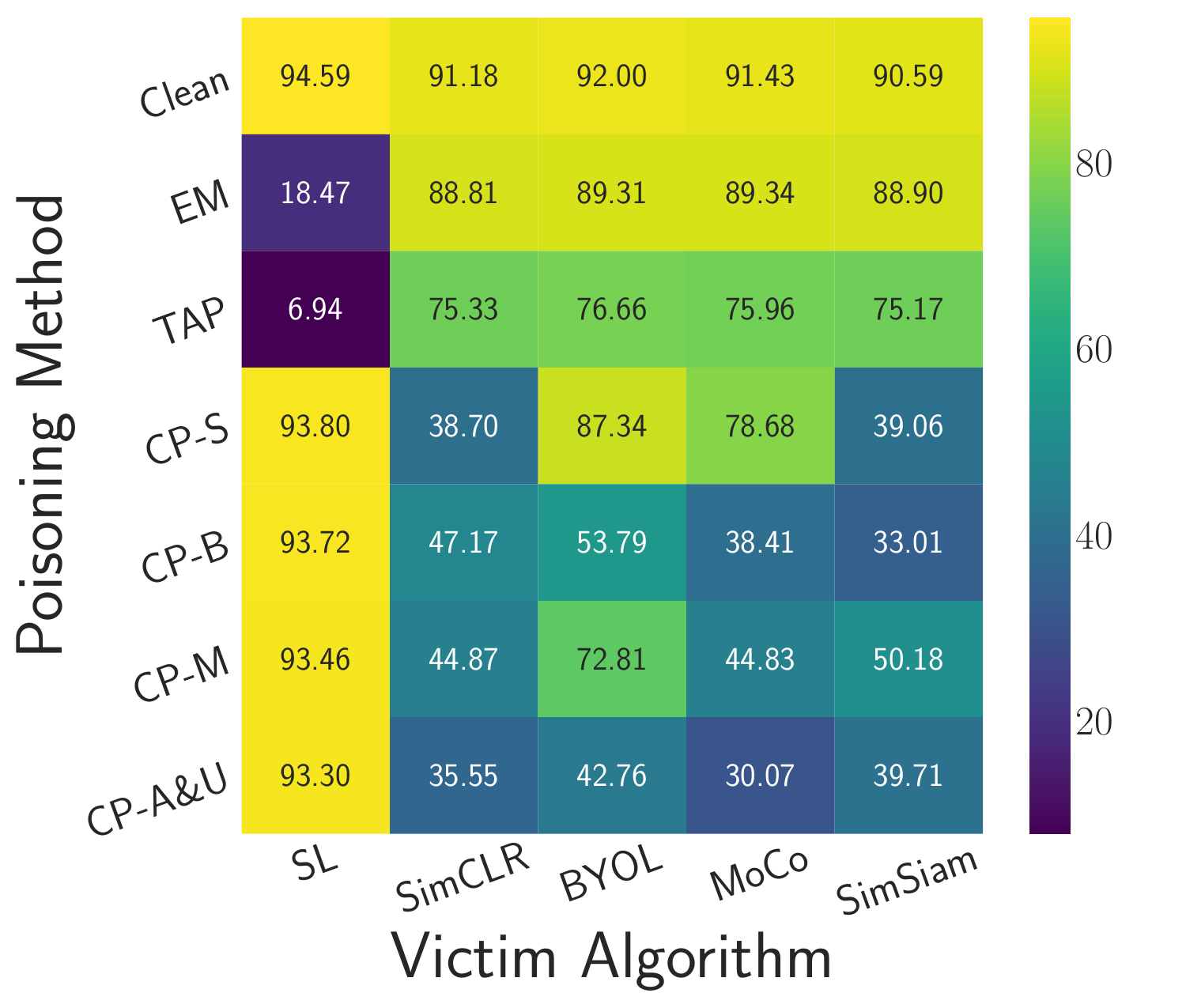}
\vspace{-12pt}
\caption{Test accuracy (\%) of victim model trained by different supervised and unsupervised contrastive learning algorithms on clean and various types of poisoned data. Here, ``CP-S'', ``CP-B'', ``CP-M'' and ``CP-A\&U'' stand for contrastive poisoning with SimCLR, BYOL, MoCo, alignment and uniformity loss, respectively. All the models are trained on ResNet-18 and CIFAR-10.}
\label{figure:motivation}
\vspace{-20pt}
\end{wrapfigure}

\subsection{Problem Setup}
\label{sec:problem}
We consider the setting where there are two parties: an adversary and a victim. The adversary is allowed to add small perturbations to any training data but has no knowledge about how the victim will train the model, including the learning paradigm, model architecture and initialization scheme, and training routine. Once the adversary releases the poisoned version of the training data, no further modification is allowed. The victim only has access to the poisoned data but with the flexibility to choose a training algorithm from various learning paradigms (supervised and contrastive learning are mainly considered) to produce a satisfactory model for the targeted task. The goal of the adversary is to downgrade the test accuracy of the trained model by crafting imperceptible perturbations to training data no matter what learning algorithm the victim adopts. We assume that the learning algorithm considered by the victim has to be competitive in that the performance of its induced model with clean training dataset is good enough.

We focus on the image classification task. Let $[n] = \{1,2,\ldots,n\}$ and $\mathcal{D}_{c} = \{(\bm{x}_i,y_i)\}_{i\in[n]}$ be the clean training dataset. The poisoned dataset is denoted as $\mathcal{D}_{p} = \{(\bm{x}_{i} + \bm{\delta}_{i},y_i)\}_{i\in[n]}$ where $\bm{\delta}_{i}$ is the poisoning perturbation added to the $i$-th training input $\bm{x}_{i}$. For the ease of presentation, we denote $\mathcal{S}_\delta = \{\bm\delta_i: i\in[n]\}$ as the set of all poisoning perturbations. To ensure the injected noise to be imperceptible to human eyes and difficult for detection, we adopt the commonly imposed $\ell_\infty$-norm constraint $\|\bm{\delta}\|_{\infty} \leq \epsilon$ for any $\bm\delta\in\mathcal{S}_\delta$, where $\epsilon>0$ denotes the predefined poison budget. Following previous works~\cite{FGCGCG21,SSGGGJ22,RXWMST23}, we focus our study on sample-wise poisons where the perturbation is generated separately for each sample. Some works~\cite{HMEBW21,HZK23} also consider class-wise poisons where all examples in the same class have the same added noise, while it has been shown that this kind of perturbations can be removed by taking average of images within the class~\cite{SSGGGJ22}.

\subsection{Proposed Method}
\label{sec:method}

\shortsection{Existing Attacks} Before presenting our \textit{Transferable Poisoning}, we first study the transferability of existing methods (Figure \ref{figure:motivation}). We test \textit{Unlearnable Examples} (EM)~\cite{HMEBW21} and \textit{Targeted Adversarial Poisoning} (TAP)~\cite{FGCGCG21} which are targeted for supervised learning, \textit{Contrastive Poisoning} (CP)~\cite{HZK23} with different frameworks which is specified for unsupervised contrastive learning. We consider that the victim can choose training algorithms from two representative learning paradigms, including standard supervised learning (SL) and four unsupervised contrastive learning algorithms SimCLR~\cite{CKNH20}, BYOL~\cite{GSATRBDPGAPKMV20}, MoCov2~\cite{CFGH20} and SimSiam~\cite{CH21}. According to Figure \ref{figure:motivation}, all of the considered learners show strong performance in producing model with high clean test accuracy, suggesting them as prominent candidate learning algorithms that the victim is likely to pick. However, Figure~\ref{figure:motivation} shows that existing availability poisoning attacks designed for a specific learning paradigm have poor transferability across various learning paradigms, especially for contrastive learning, where the poisoning effect cannot transfer within the same learning paradigm while with different frameworks.

\begin{algorithm}[t] 
\caption{Transferable Poisoning} 
\label{alg:tp}
\begin{algorithmic}[1]
\State {\bfseries Input:} clean dataset $\mathcal{D}_c$; number of total epochs $T$, number of updates in each epoch $M$; learning rate $\eta_{\mathrm{sl}}$, $\eta_{\mathrm{cl}}$; PGD steps $S_{\mathrm{sl}}$, $S_{\mathrm{cl}}$; attack step size $\alpha_{\mathrm{sl}}$, $\alpha_{\mathrm{cl}}$; poison budget $\epsilon$
\For {$t = 1,\ldots,T$}
    \For {$m$ in $1, \ldots,M$}
        \State Sample $\{\bm{x}_i,y_{i}\}_{i\in[B]}$ from $\mathcal{D}_c$
        \State  $\theta_b+\theta_c \leftarrow \theta_b+\theta_c - \eta_\mathrm{cl} \cdot \nabla_{\theta_b+\theta_c} \sum_{i\in[B]}\mathcal{L}_{\mathrm{CL}}(f(\{\bm{x}^{a}_i + \bm{\delta}_i\},\{\bm{x}^{b}_i + \bm{\delta}_i\};\theta_b+\theta_c))$ 
    \EndFor
    \For {$m$ in $1, \ldots,M$}
        \State Sample $\{\bm{x}_i,y_{i}\}_{i\in[B]}$ from $\mathcal{D}_c$
                \For {$s$ in $1, \ldots,S_{\mathrm{cl}}$}
                    \State Compute $\bm{g}_i \leftarrow \nabla_{\bm{\delta}_i} \sum_{i\in[B]}\mathcal{L}_{\mathrm{CL}}(f(\{\bm{x}^{a}_i + \bm{\delta}_i\},\{\bm{x}^{b}_i + \bm{\delta}_i\};\theta_b+\theta_c)),  \forall i\in[B]$
                    \State $\bm{\delta}_i \leftarrow \Pi_{\mathcal{B}_\epsilon(\bm{0)}} \left( \bm{\delta}_i - \alpha_{\mathrm{cl}} \cdot \mathrm{sgn}(\bm{g}_i) \right),  \forall i\in[B]$ 
        \EndFor
    \EndFor
    \For {$m$ in $1, \ldots,M$}
        \State Sample $\{\bm{x}_i,y_{i}\}_{i\in[B]}$ from $\mathcal{D}_c$
        \State  $\theta_b+\theta_s \leftarrow \theta_b+\theta_s - \eta_\mathrm{sl} \cdot \nabla_{\theta_b+\theta_s} \sum_{i\in[B]}\mathcal{L}_{\mathrm{CE}}(f(\{\bm{x}_i + \bm{\delta}_i\};\theta_b+\theta_s))$ 
    \EndFor
    \For {$m$ in $1, \ldots,M$}
        \State Sample $\{\bm{x}_i,y_{i}\}_{i\in[B]}$ from $\mathcal{D}_c$
                \For {$s$ in $1, \ldots,S_{\mathrm{sl}}$}
                    \State Compute $\bm{g}_i \leftarrow \nabla_{\bm{\delta}_i} \sum_{i\in[B]}\mathcal{L}_{\mathrm{CE}}(f(\{\bm{x}_i + \bm{\delta}_i\};\theta_b+\theta_s)),  \forall i\in[B]$
                    \State $\bm{\delta}_i \leftarrow \Pi_{\mathcal{B}_\epsilon(\bm{0)}} \left( \bm{\delta}_i - \alpha_{\mathrm{sl}} \cdot \mathrm{sgn}(\bm{g}_i) \right),  \forall i\in[B]$ 
        \EndFor
    \EndFor
\EndFor
\State {\bfseries Output:} poisoned dataset $\mathcal{D}_{p} = \{(\bm{x}_i + \bm{\delta}_i,y_i)\}_{i\in[n]}$
\end{algorithmic} 
\end{algorithm}

\shortsection{Transferable Poisoning}
Similar to~\cite{WZG24}, we consider the \emph{worst-case unlearnability} across a set of supervised and contrastive learning algorithms $\{\mathcal{A}_1, \mathcal{A}_2, \ldots, \mathcal{A}_K\}$. More rigorously, it can be cast as the following optimization problem:
\begin{equation}
    \min_{\mathcal{S}_\delta}\max_{k\in [K]}\EE_{(\bm{x},y)\sim\mathcal{D}}\big[\textbf{1}(f_{\theta_k}(\bm{x}) = y)\big] \:\text{ s.t. }\:
    \theta_k = \argmin_{\theta} \frac{1}{n}\sum_{i=1}^n \mathcal{L}_{k}(f(\bm{x}_i + \bm{\delta}_i),y_i; \mathcal{A}_k).
\end{equation}
where $\mathcal{D}$ denotes the underlying clean data distribution, and $\mathcal{L}_{k}$ is the training loss corresponded to $\mathcal{A}_k$. It is expected to generate the poisons according the model with the highest accuracy, and thus achieve the worst-case unlearnability. However, direct optimization is not feasible since it is difficult to enumerate all training algorithms, especially for contrastive learning, and how to make the effect of the poisons generated based on one learning algorithm transfer to the other is another problem. To solve these difficulties, we adopt two essential approximations.

First, as shown in Figure \ref{figure:motivation}, the poisoning perturbations generated based on one specific contrastive learning framework is difficult to transfer even within the same learning paradigm, including representative training algorithms such as SimCLR, BYOL and MoCov2. This observation motivates us to exploit the intrinsic characteristics for poisoning contrastive learning. Generally speaking, contrastive loss learns representations by pulling together the anchor image and a positive sample, while pushing apart the anchor and the other negative samples in the feature space. Previous works~\cite{OLV18,TKI202,HFLGBTB19} argue that representations should be invariant to other noise factors and preserve as much information as possible, which are denoted as alignment and uniformity~\cite{WI20}:
\begin{equation}
    \mathcal{L}_{\mathrm{align}}(f, \bm{x}^{a}_{i}, \bm{x}^{b}_{i}; \alpha) = \E\big[||f(\bm{x}^{a}_{i})-f(\bm{x}^{b}_{i})||^{\alpha}_{2}\big] \:\: \text{and} \:\:
    \mathcal{L}_{\mathrm{uniform}}(f, \bm{x}; t) = \log\E\big[e^{-t||f(\bm{x})||^2_2}\big],
\end{equation}
where $\bm{x}^{a}_{i}$ and $\bm{x}^{b}_{i}$ are the two augmented views of the same data point $\bm{x}_i$, $\alpha$ and $t$ are two hyperparameters for alignment and uniformity. Then the final loss is the weighted combination of them:
\begin{equation}
    \mathcal{L}_{\mathrm{CL}}(f, \bm{x}^{a}_{i}, \bm{x}^{b}_{i}) = \lambda\mathcal{L}_{\mathrm{align}}(f, \bm{x}^{a}_{i}, \bm{x}^{b}_{i}, \alpha) + (\mathcal{L}_{\mathrm{uniform}}(f, \bm{x}^{a}_{i}, t) + \mathcal{L}_{\mathrm{uniform}}(f, \bm{x}^{b}_{i}, t))/2.
\end{equation}
with $\lambda$ as the factor to balance these two losses. Wang et al.~\cite{WI20} demonstrate strong connection of contrastive loss with alignment and uniformity through both theoretical analysis and extensive experiments. Inspired by this, we propose to leverage alignment and uniformity to generate contrastive poisons:
\begin{equation}
    \min_{\theta, \mathcal{S}_\delta} \:\frac{1}{n}\sum_{j\in[N]}\sum_{i\in[B]} \mathcal{L}_{\mathrm{CL}}\big(f(\{\bm{x^a_i} + \bm{\delta}_i\}, \{\bm{x^b_i} + \bm{\delta}_i\};\theta)\big), \text{ s.t. } \|\bm\delta_i\|_\infty \leq \eps,  \forall i\in[n].
\end{equation}
where $N$ denotes the number of batches, each batch has the same size of $B$ training data with $n=N\cdot B$. The above optimization is solved by alternating between updating the model and poisoning perturbation. As can be found in Figure \ref{figure:motivation}, the poisons generated with alignment and uniformity loss indeed have better transferability in contrastive learning.

Second, in order to make the poisoning perturbation aligned with supervised and contrastive learning, we propose a simple training strategy, that is, we iteratively update the noise based on the model trained by supervised (cross-entropy loss) and contrastive learning (alignment and uniformity loss). Since the poisons for contrastive learning is essentially an error-minimization noise, here we adopt EM, the same type of noise for poisoning supervised learning:
\begin{equation}
    \min_{\theta, \mathcal{S}_\delta} \:\frac{1}{n}\sum_{i\in[n]} \mathcal{L}_{\mathrm{CE}}\big(f(\{\bm{x_i} + \bm{\delta}_i\};\theta)\big), \text{ s.t. } \|\bm\delta_i\|_\infty \leq \eps,  \forall i\in[n].
\end{equation}
The general idea is, when the model is updated based on one training algorithm, then this updated model tends to have the highest accuracy on the poisoned data, thus optimizing the poisons based on the same model is an approximated way for the poisons to achieve the worst-case unlearnability. Actually, to make the optimization direction more aligned with different training algorithms, we use a shared model during training. Since the model architectures used in supervised and contrastive learning have a little bit difference on the last linear layer, we denote them as $\theta_b+\theta_s$ and $\theta_b+\theta_c$, with $\theta_b$ as the shared backbone and $\theta_s$ and $\theta_c$ as the last linear layer for supervised and contrastive learning respectively. And we also show the advantage of choosing error-minimization for poisoning both supervised and contrastive learning compared to other alternative ways (See Section~\ref{sec:discussion} for more explanations).

\section{Experiments}
\label{sec:experiment}

\shortsection{Experimental Settings}
\label{sec:setting}
We adopt four commonly used benchmark datasets in image classification task: CIFAR-10, CIFAR-100~\cite{krizhevsky2009learning}, TinyImageNet~\cite{CLH17} and MiniImageNet~\cite{VBLKW16}. The poisoning perturbations are generated by PGD~\cite{MMSTV18} on ResNet-18~\cite{HZRS16} and we use alignment and uniformity for unsupervised contrastive learning. Following previous works, we set $\epsilon = 8/255$ to constrain the poisoning perturbations (additional results for other perturbation strength are provided in Table \ref{table:other_poisoning_budgets} in Appendix \ref{sec:additional_results}). 

Apart from the five training algorithms adopted in Section~\ref{sec:method}, we also involve supervised contrastive learning~\cite{KTWSTIMLK20} to evaluate the worst-case unlearnability i.e., the highest accuracy obtained across selected learning algorithms. For baseline methods, we select TAP, EM, CP, TUE, AUE and AAP. Specifically, the generation of TUE attacks is based on SimCLR framework. AAP has targeted and untargeted versions (T-AAP and UT-AAP), while we only generate UT-AAP on CIFAR-10 since the generation of untargeted adversarial poisoning is unstable as claimed in the original paper~\cite{WZG24}. More detailed experimental settings can be found in Appendix~\ref{sec:detailed_settings}.

\begin{wrapfigure}{R}{0.4\textwidth}
\hspace{10pt}
\includegraphics[width=0.35\textwidth]{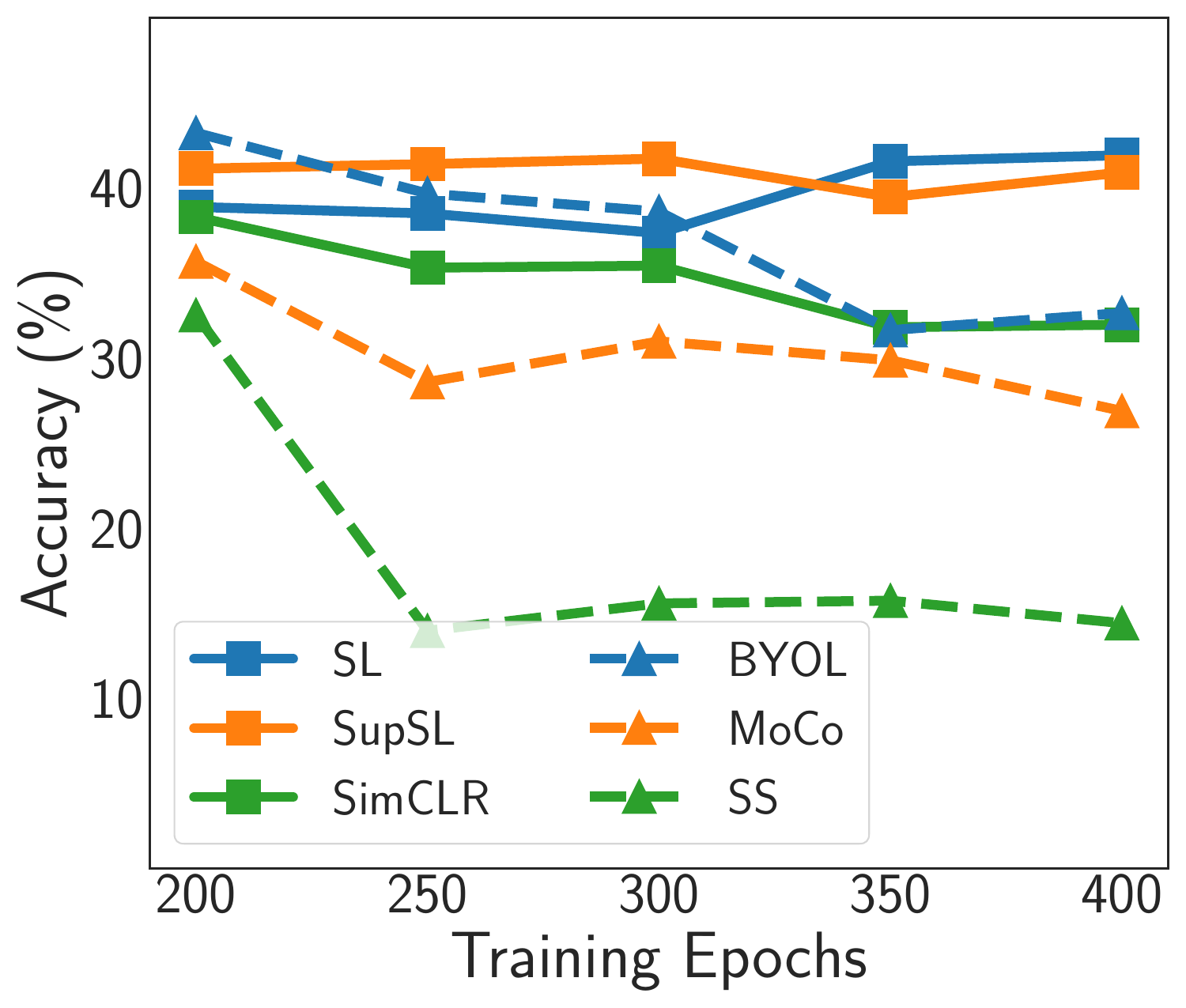}
\caption{Effect of training epochs for generating poisons on CIFAR-10.}
\label{figure:training_epochs}
\end{wrapfigure}

\begin{table*}[t]
\centering
\caption{Results of transferability across various learning paradigms and algorithms between different attacking methods. The results are clean accuracy (\%) tested on CIFAR-10.}
\vspace{-2mm}
\setlength{\tabcolsep}{4.0pt}
\label{table:main_results}
\begin{tabular}{l|c|ccccc|c}
\toprule
\rowcolor{white}
Attack & SL & SupCL & SimCLR & BYOL & MoCo & SimSiam & Worst\\
\midrule
Clean & 94.59 & 94.75 & 91.18 & 92.00 & 91.43 & 90.59 & 94.75 \\
EM & 18.47 & 89.04 & 88.81 & 89.31 & 89.34 & 88.90 & 89.34 \\
TAP & 6.94 & 60.62 & 75.33 & 76.66 & 75.96 & 75.17 & 76.66 \\
CP-SimCLR & 93.80 & 94.21 & 38.70 & 87.34 & 78.68 & 39.06 & 94.21 \\
CP-BYOL & 93.72 & 94.07 & 47.17 & 53.79 & 38.41 & 33.01 & 94.07 \\
CP-MoCo & 93.46 & 93.94 & 44.87 & 72.81 & 44.83 & 50.18 & 93.94 \\
TUE & 29.37 & 84.22 & 47.46 & 88.31 & 83.02 & 46.96 & 88.31 \\ 
AUE & 22.15 & 32.79 & 58.97 & 64.79 & 62.57 & 27.03 & 64.79 \\
T-AAP & 10.44 & 25.19 & 42.94 & 48.61 & 48.58 & 45.03 & 48.61 \\
UT-AAP & 35.60 & 54.53 & 37.99 & 41.58 & 37.80 & 39.34 & 54.53 \\ 
\midrule
Ours & 37.34 & 41.72 & 35.43 & 38.61 & 30.98 & 15.58 & 41.72 \\
\bottomrule
\end{tabular}
\end{table*}

\begin{table*}[t]
\centering
\caption{Results of transferability across various learning paradigms and algorithms between different attacking methods. The results are clean accuracy (\%) tested on CIFAR-100.}
\vspace{-2mm}
\setlength{\tabcolsep}{4.0pt}
\label{table:main_results_cifar100}
\begin{tabular}{l|c|ccccc|c}
\toprule
\rowcolor{white}
Attack & SL & SupCL & SimCLR & BYOL & MoCo & SimSiam & Worst\\
\midrule
Clean & 75.55 & 73.17 & 63.52 & 65.03 & 64.46 & 62.45 & 75.55 \\
EM & 15.57 & 69.07 & 55.26 & 63.21 & 59.81 & 52.81 & 69.07 \\
TAP & 3.94 & 25.23 & 26.15 & 26.55 & 28.97 & 28.84 & 28.97 \\
CP-SimCLR & 73.24 & 66.46 & 8.62 & 39.41 & 35.67 & 7.17 & 73.24 \\
CP-BYOL & 73.69 & 67.35 & 10.22 & 16.90 & 7.27 & 10.30 & 73.69 \\
CP-MoCo & 73.35 & 67.12 & 8.20 & 10.97 & 5.64 & 4.69 & 73.35  \\
TUE & 1.34 & 68.54 & 55.41 & 55.09 & 51.68 & 53.04 & 68.54 \\
AUE & 3.37 & 57.28 & 61.11 & 63.20 & 62.44 & 60.79 & 63.20 \\
T-AAP & 7.17 & 18.16 & 18.31 & 21.37 & 20.26 & 20.89 & 21.37 \\
\midrule
Ours & 9.54 & 14.25 & 14.81 & 16.51 & 15.37 & 15.25 & 16.51 \\
\bottomrule
\end{tabular}
\end{table*}

\begin{table*}[t]
\centering
\caption{Results of transferability across various learning paradigms and algorithms between different attacking methods. The results are clean accuracy (\%) tested on TinyImageNet and MiniImageNet.}
\vspace{-2mm}
\setlength{\tabcolsep}{4.0pt}
\label{table:main_results_tiny_mini}
\begin{tabular}{ll|c|ccccc|c}
\toprule
\rowcolor{white}
Dataset & Attack & SL & SupCL & SimCLR & BYOL & MoCo & SimSiam & Worst\\
\midrule
& Clean & 63.07 & 59.51 & 48.47 & 41.16 & 45.90 & 32.85 & 63.07 \\
& TAP & 17.55 & 35.93 & 37.51 & 27.42 & 36.68 & 29.80 & 37.51 \\
TinyImageNet & AUE & 4.57 & 15.43 & 26.77 & 28.51 & 22.20 & 26.58 & 28.51 \\ 
& T-AAP & 18.84 & 25.63 & 26.03 & 27.95 & 24.55 & 23.53 & 27.95 \\
& Ours & 3.45 & 3.60 & 3.43 & 3.78 & 4.57 & 2.83 & 4.57 \\
\midrule
& Clean & 75.35 & 70.82 & 68.03 & 66.22 & 70.20 & 58.15 & 75.35 \\
& TAP & 20.21 & 42.43 & 46.90 & 46.44 & 46.34 & 39.34 & 46.90 \\
MiniImageNet & AUE & 5.35 & 47.52 & 54.22 & 56.78 & 56.62 & 49.77 & 56.78 \\
& T-AAP & 21.83 & 27.70 & 27.23 & 33.03 & 26.14 & 26.66 & 33.03 \\
& Ours & 4.62 & 6.47 & 6.63 & 7.32 & 9.34 & 6.12 & 9.34 \\
\bottomrule 
\end{tabular}
\end{table*}

\subsection{Comparisons with Existing Attacks}
\label{sec:main_results}
We first evaluate the transferability of our method on the benchmark CIFAR-10 and CIFAR-100 and the results are in Table~\ref{table:main_results} and~\ref{table:main_results_cifar100}. Most existing methods struggle to achieve satisfactory transferability to learning algorithms that are not directly used in the perturbation generation process. More specifically, EM is originally designed for supervised learning while exhibits poor performance in contrastive learning. CP has minimal effect on learning with supervision including SL and SupCL, and even can't transfer well within unsupervised contrastive learning for some frameworks such as SimCLR. TUE shows relatively improved transferability, while supervised contrastive learning can largely restore the model's accuracy. AUE and UT-AAP on CIFAR-10, TAP on CIFAR-100, and T-AAP on both CIFAR-10 and CIFAR-100 are the comparatively successful attacks in terms of the worst-case unlearnability, while as a comparison, our attack still achieves state-of-the-art transferability. We then compare our attack with these successful baselines on high-resolution datasets in Table~\ref{table:main_results_tiny_mini}. We find our attack have more advantages on these large-scale datasets, as we improve the results by 23.38\% and 23.69\% over the baselines for TinyImageNet and MiniImageNet respectively. In general, contrastive learning is more robust to availability attacks than supervised learning, as the accuracy recovered by the former is much higher than the latter in the attacks that considers the transferability such as TUE, AUE and T-AAP. While our attack not only improves the transferability within contrastive learning to a large extent, but also finds a better way to make the poisons transfer between supervised and contrastive learning.

\begin{table*}[t]
\centering
\caption{Effect of weight between alignment and uniformity.}
\vspace{-2mm}
\setlength{\tabcolsep}{4.0pt}
\label{table:align_uniform_weight}
\begin{tabular}{l|c|ccccc|c}
\toprule
\rowcolor{white}
Weight & SL & SupCL & SimCLR & BYOL & MoCo & SimSiam & Worst\\
\midrule
3:1 & 35.90 & 38.29 & 33.95 & 39.29 & 33.68 & 16.99 & 39.29 \\
2:1 & 38.57 & 38.71 & 35.66 & 38.37 & 34.00 & 18.19 & 38.71  \\
1:1 & 37.34 & 41.72 & 35.43 & 38.61 & 30.98 & 15.58 & 41.72 \\
1:2 & 43.88 & 40.24 & 34.04 & 33.45 & 30.33 & 12.79 & 43.88 \\
1:3 & 46.02 & 41.78 & 36.40 & 37.48 & 29.05 & 11.78 & 46.02 \\
\bottomrule
\end{tabular}
\end{table*}

\subsection{Effect of Weight between Alignment and Uniformity}
\label{sec:align_uniform_weight}
Since our attack proposes to leverage alignment and uniformity loss to generate contrastive poisons, the weight between these two terms is an essential factor for the attack performance. As shown in Table~\ref{table:align_uniform_weight}, assigning higher weight on alignment loss will comparatively improve the unlearnability of poisons on SL, SupSL and SimCLR, while influence the attack performance on BYOL, MoCo and SimSiam in a negative way. In our default settings, we set the same weight on alignment and uniformity for all the datasets, and it achieves good attack performance. While tuning this hyper-parameter can enable even better results in terms of the worst-case unlearnability, e.g., using slightly higher weight on alignment loss for CIFAR-10 dataset.

\begin{table*}[t]
\centering
\caption{Effect of ratio of PGD steps for generating poisons on CIFAR-10.}
\vspace{-2mm}
\setlength{\tabcolsep}{4.0pt}
\label{table:pgd_steps}
\begin{tabular}{l|c|ccccc|c}
\toprule
\rowcolor{white}
Ratio & SL & SupCL & SimCLR & BYOL & MoCo & SimSiam & Worst\\
\midrule
3:3 & 41.02 & 41.54 & 34.77 & 38.27 & 30.14 & 20.78 & 41.54  \\
5:5 & 37.34 & 41.72 & 35.43 & 38.61 & 30.98 & 15.58 & 41.72 \\
7:7 & 41.12 & 42.24 & 35.72 & 35.75 & 31.25 & 13.37 & 42.24 \\
9:9 & 42.46 & 40.63 & 34.55 & 34.99 & 30.72 & 14.32 & 42.46 \\
\midrule
3:5 & 38.37 & 44.62 & 45.64 & 48.11 & 46.59 & 38.34 & 48.11 \\
5:3 & 40.12 & 43.60 & 38.59 & 41.27 & 37.91 & 17.31 & 43.60 \\
\bottomrule
\end{tabular}
\end{table*}

\begin{table*}[t]
\centering
\caption{Transferability under model architectures on CIFAR-10.}
\vspace{-2mm}
\setlength{\tabcolsep}{4.0pt}
\label{table:model_architecture}
\begin{tabular}{l|c|ccccc|c}
\toprule
\rowcolor{white}
Architecture & SL & SupCL & SimCLR & BYOL & MoCo & SimSiam & Worst\\
\midrule
ResNet-18 & 37.34 & 41.72 & 35.43 & 38.61 & 30.98 & 15.58 & 41.72 \\
ResNet-34 & 38.99 & 42.60 & 34.38 & 39.24 & 33.73 & 31.97 & 42.60 \\
VGG-19 & 39.82 & 44.81 & 40.43 & 31.44 & 38.29 & 10.12 & 44.81 \\
DenseNet-121 & 39.43 & 44.93 & 38.77 & 43.05 & 37.23 & 40.30 & 44.93 \\
MobileNetV2 & 38.63 & 39.42 & 37.72 & 42.82 & 34.33 & 10.89 & 41.82 \\
\bottomrule
\end{tabular}
\end{table*}

\subsection{Effect of Training Epochs and Iteration Ratios}
\label{sec:iteration_ratio}
In this section, we change the training epochs and ratio of PDG steps for poisoning supervised and contrastive learning to see how they influence the poisoning effect. As shown in Figure~\ref{figure:training_epochs}, the poisons will generally have better unlearnability on unsupervised contrastive learning as the training epochs increase, while the attack performance on SL will be slightly downgraded. It means poisoning contrastive learning needs more training epochs than SL, however, we can obtain desirable worst-case unlearnability when the training epochs are in an appropriate range. When it comes to the ratio of PGD steps, Table~\ref{table:pgd_steps} demonstrates that the unlearnability of poisons on all the training algorithms are stable when we vary the PGD steps for supervised and contrastive learning while keep the ratio to be $1$. However, the unlearnability on contrastive learning will be highly influenced when the ratio is not balanced, especially if we use less PGD steps for contrastive learning. Interestingly, the attack performance on contrastive learning will still be downgraded even we use more PGD steps on it as the case of $5:3$ compared to the default setting $5:5$. It can be explained that for contrastive learning, the final linear probing is actually supervised learning, so more poisoning effect on supervised learning will improve the total unlearnability. In general, when we use a balanced ratio of PGD steps, less training epochs or optimization steps will still give us desirable attack performance.

\subsection{Transferability under Model Architectures}
\label{sec:model_architecture}
In default settings, we use ResNet-18 as the architecture for both supervised and unsupervised contrastive learning to generate the poisoning perturbations, here we test whether such poisons are still effective if the victim adopts other architectures to learn a model. As can be seen from Table~\ref{table:model_architecture}, our attack can transfer quite well from ResNet-18 to other five commonly used model architectures including ResNet-34, VGG-19, DenseNet-121 and MobileNetV2, which further demonstrates the strong transferability of our proposed method.

\subsection{Effect of Partial Poisoning}
\label{sec:partial_poison}
\begin{figure*}[!t]
\centering
\captionsetup[subfigure]{justification=centering}
\begin{subfigure}{0.14\columnwidth}
\includegraphics[width=\columnwidth]{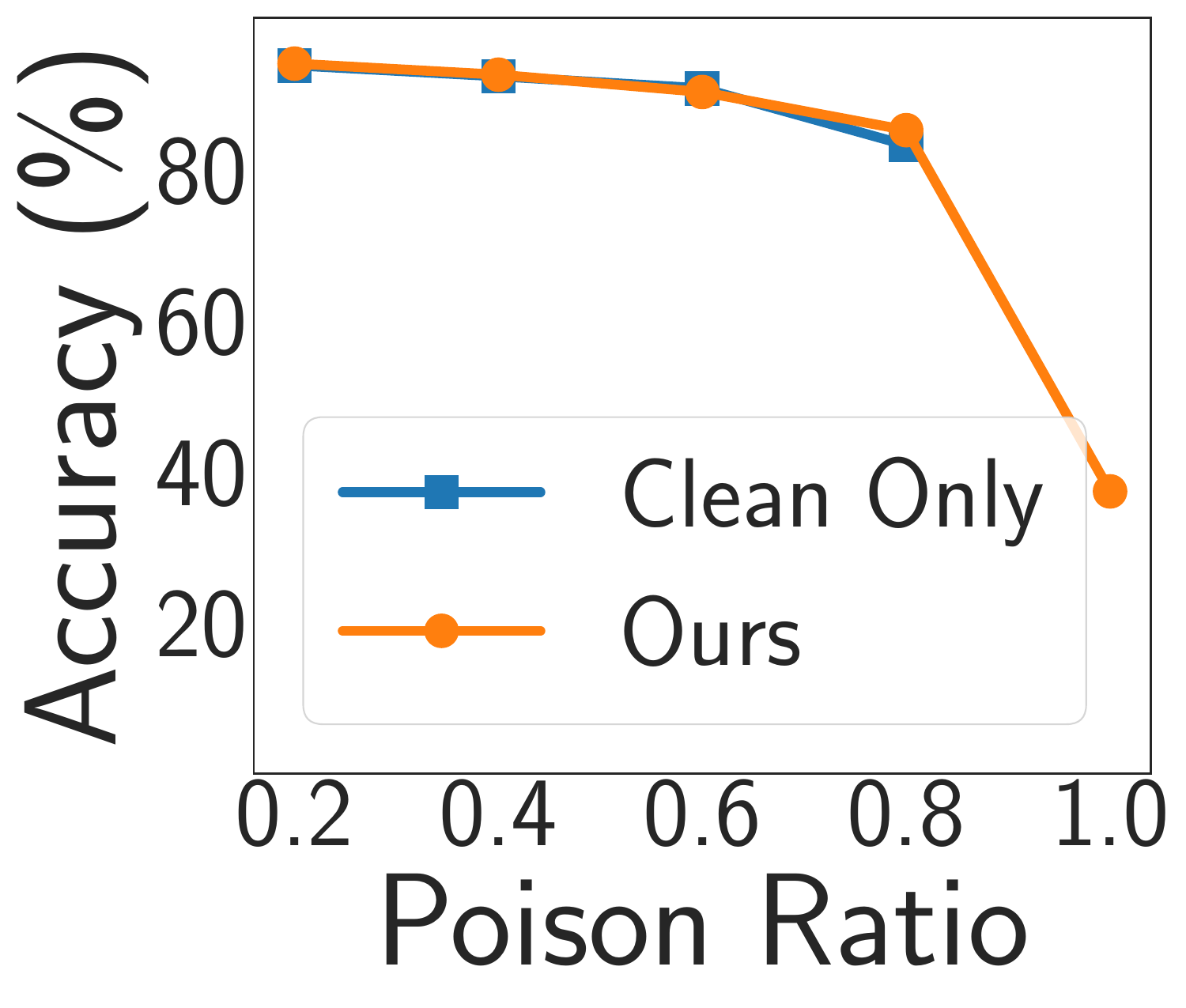}
\caption{SL}
\label{figure:partial_supervised}
\end{subfigure}
\hspace{2mm}
\begin{subfigure}{0.14\columnwidth}
\includegraphics[width=\columnwidth]{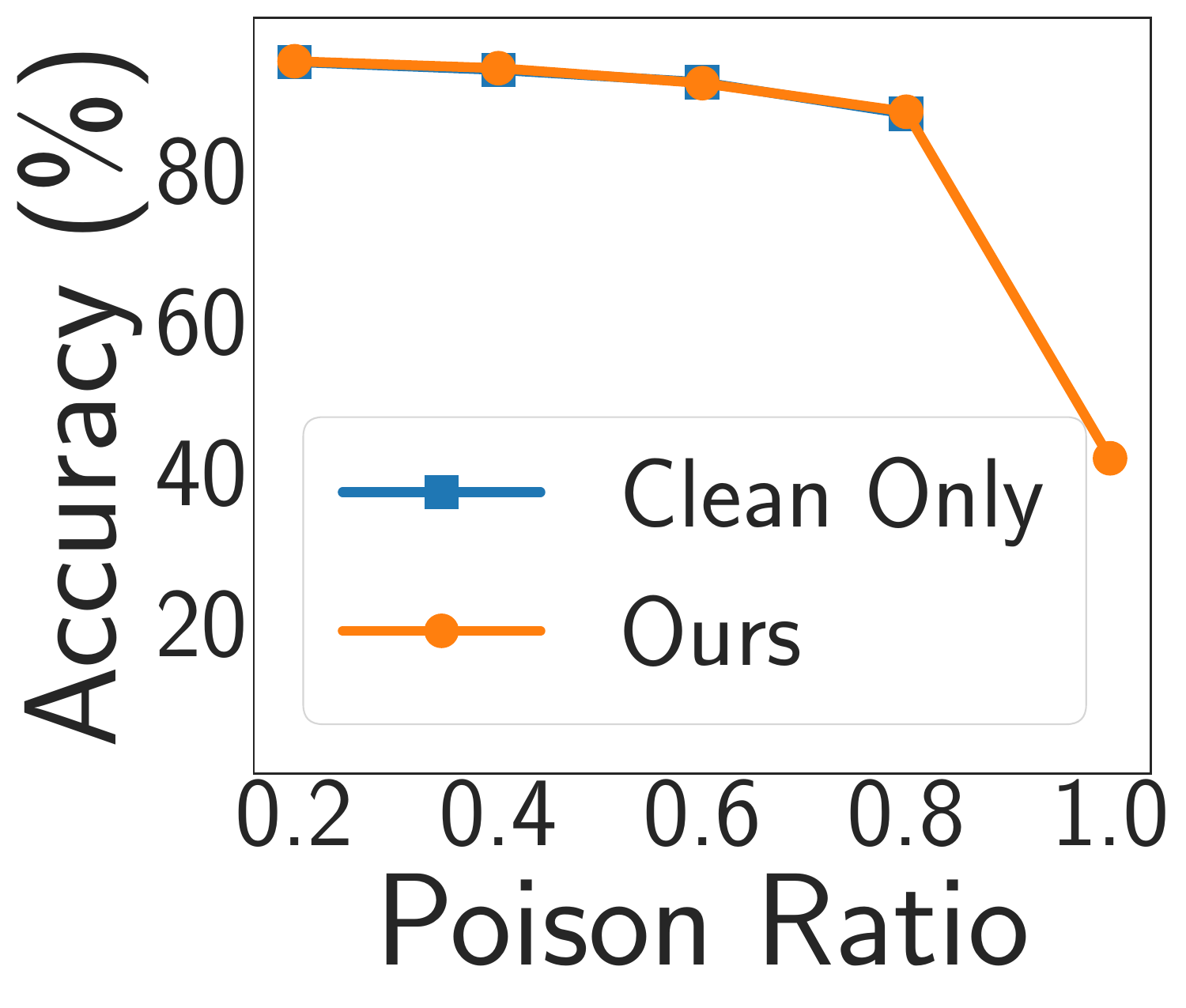}
\caption{SupCL}
\label{figure:partial_scl}
\end{subfigure}
\hspace{2mm}
\begin{subfigure}{0.14\columnwidth}
\includegraphics[width=\columnwidth]{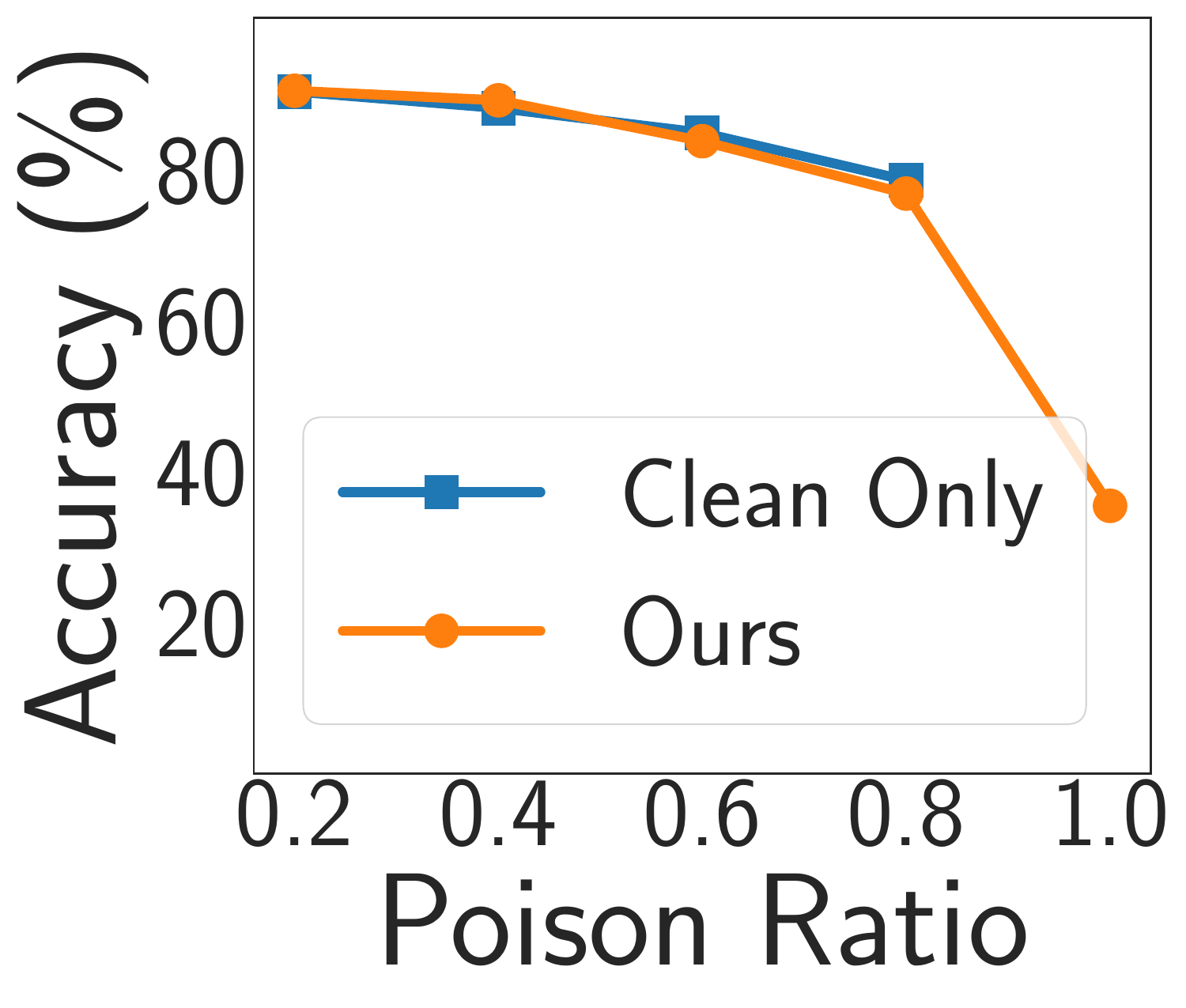}
\caption{SimCLR}
\label{figure:partial_simclr}
\end{subfigure}
\hspace{2mm}
\begin{subfigure}{0.14\columnwidth}
\includegraphics[width=\columnwidth]{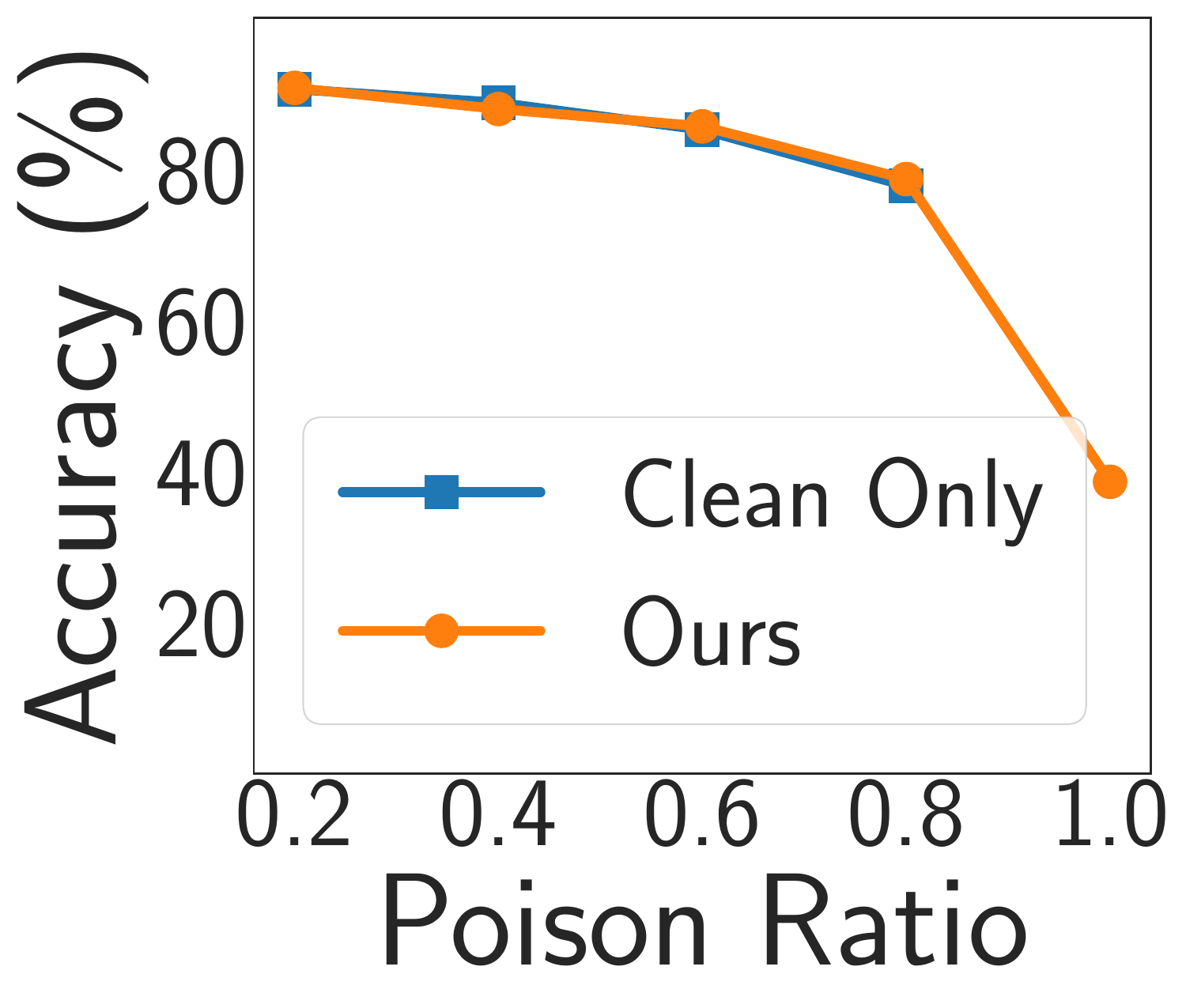}
\caption{BYOL}
\label{figure:partial_byol}
\end{subfigure}
\hspace{2mm}
\begin{subfigure}{0.14\columnwidth}
\includegraphics[width=\columnwidth]{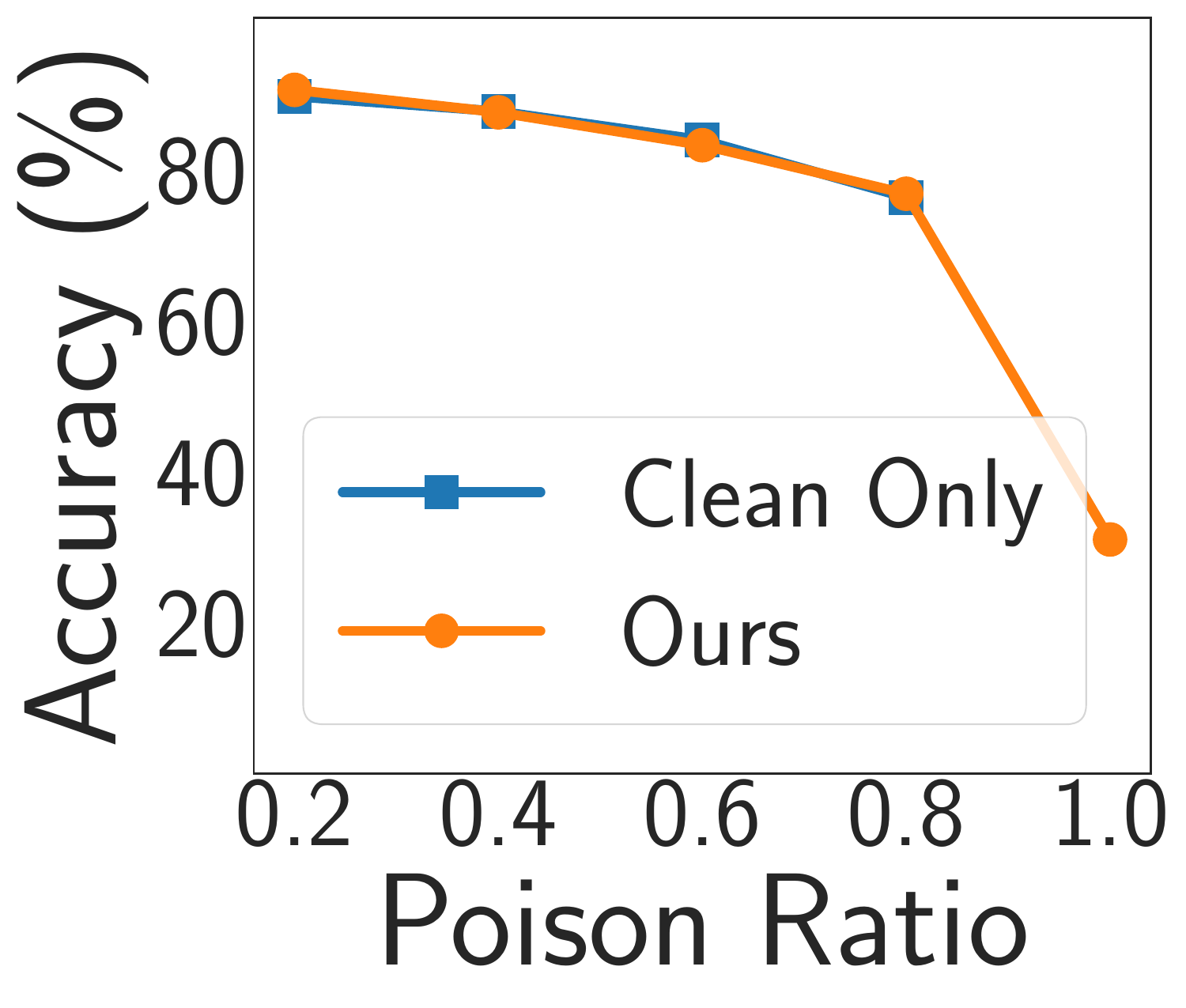}
\caption{MoCo}
\label{figure:partial_moco}
\end{subfigure}
\hspace{2mm}
\begin{subfigure}{0.14\columnwidth}
\includegraphics[width=\columnwidth]{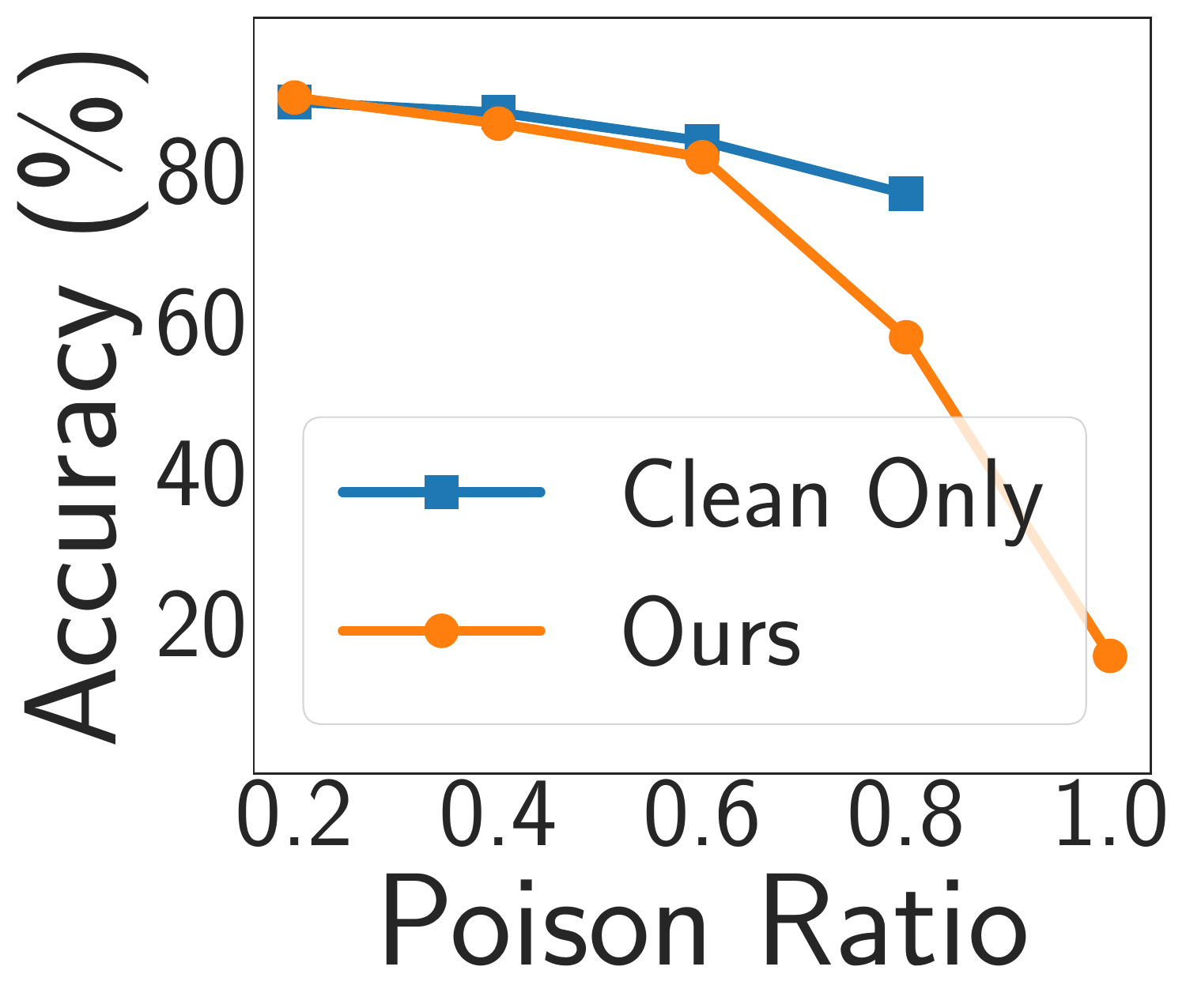}
\caption{SimSiam}
\label{figure:partial_SS}
\end{subfigure}
\caption{Effect of partial poisoning on CIFAR-10. ``Ours'' uses the entire training data with the poisoning proportion adjusted, while ``Clean Only'' merely uses the rest clean training data.}
\label{figure:partial_poison}
\end{figure*}

Here, we consider more challenging settings where only partial training data can be poisoned. We report the results in Figure~\ref{figure:partial_poison}. As for the reference where the model is trained with the rest clean data, the test accuracy shows desirable patterns for all the selected training algorithms as it gradually decreases when the poison ratio increases. While the poisoning effect is vulnerable to the ratio of poisoning as the accuracy of the poisoned model will sharply increases when the poison proportion is not 100\%, which shows the same characteristics as the previous attacks~\cite{HMEBW21,FGCGCG21,HZK23,RXWMST23,WZG24}.

\begin{table*}[t]
\centering
\caption{Alternative attacking methods on CIFAR-10.}
\vspace{-2mm}
\setlength{\tabcolsep}{4.0pt}
\label{table:different_attacks}
\begin{tabular}{l|c|ccccc|c}
\toprule
\rowcolor{white}
Attack & SL & SupCL & SimCLR & BYOL & MoCo & SimSiam & Worst\\
\midrule
Clean & 94.59 & 94.75 & 91.18 & 92.00 & 91.43 & 90.59 & 94.75 \\
HF & 16.00 & 92.12 & 63.68 & 91.65 & 83.75 & 63.47 & 92.12 \\
CC & 16.50 & 90.27 & 47.31 & 89.06 & 41.16 & 43.31 & 90.27 \\
Two models & 17.73 & 84.11 & 50.48 & 58.16 & 47.36 & 51.19 & 84.11 \\
TAP-based  & 5.97 & 32.00 & 56.52 & 55.87 & 41.17 & 48.28 & 56.52 \\
Ours & 37.34 & 41.72 & 35.43 & 38.61 & 30.98 & 15.58 & 41.72 \\
\bottomrule
\end{tabular}
\end{table*}

\section{Discussion}
\label{sec:discussion}
To examine the effectiveness of our \textit{Transferable Poisoning} in leveraging the information from supervised and unsupervised contrastive learning, we first compare our method with two naive alternative ways. One is to directly add the poisons generated by EM and CP with alignment and uniformity separately, while the poison budget for each is set as $\epsilon/2$, denoted as HF, to ensure the final poisons are under the budget $\epsilon$. The other is created by first combining the two poisons which are generated under the full poison budget $\epsilon$, and then clamping it to meet the constraint, which is denoted as CC. We also substitute some components of our method to form another two attacks. In our TP, we share the backbone of the model for poisoning supervised and unsupervised contrastive learning, while here we try to use two models separately in the iterative process. We also change the method for poisoning SL from EM to TAP to form ``Tap-based'' attack. Since TAP and CP work differently as the former needs to pretrain the model first and then generate the poisons while the latter trains the model and generates the poisons iteratively, we have to use two models for this attack.

Table~\ref{table:different_attacks} summaries the results. All of them can keep the poisoning effect on supervised learning, while more or less downgrade the performance on contrastive learning. HF, CC and poisoning with two models have negligible unlearnability on supervised contrastive learning and even BYOL for HF and CC. Among them, TAP-based method inherits the merits for poisoning supervised learning, and has less negative effect on contrastive learning. While comparing to our \textit{Transferable Poisoning}, it still influences the attack performance on unsupervised contrastive learning to a large extent. 

\begin{figure*}[!t]
\centering
\begin{subfigure}{0.14\columnwidth}
\includegraphics[width=\columnwidth]{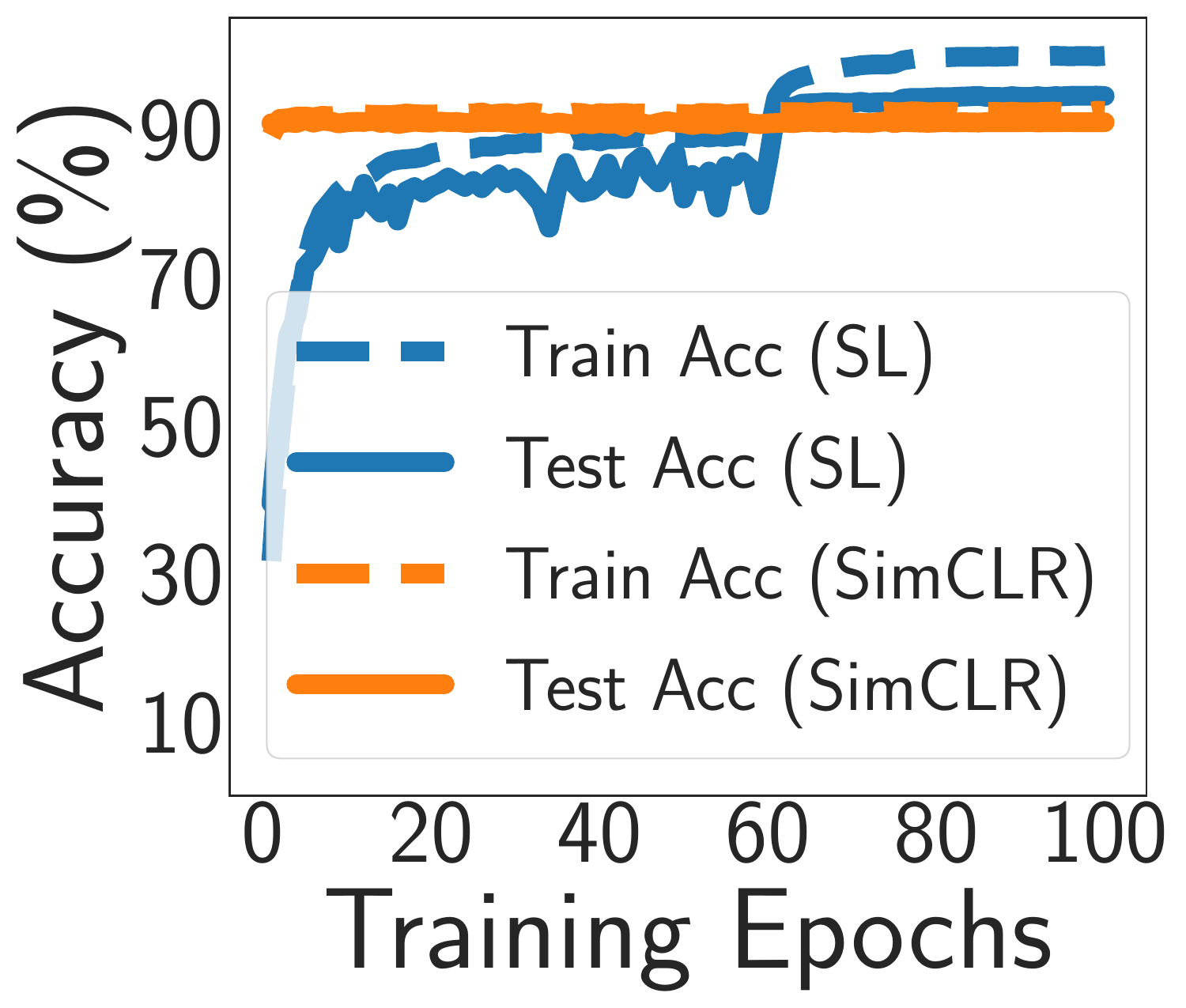}
\caption{Clean}
\label{figure:different_training_methods_clean}
\end{subfigure}
\hspace{2mm}
\begin{subfigure}{0.14\columnwidth}
\includegraphics[width=\columnwidth]{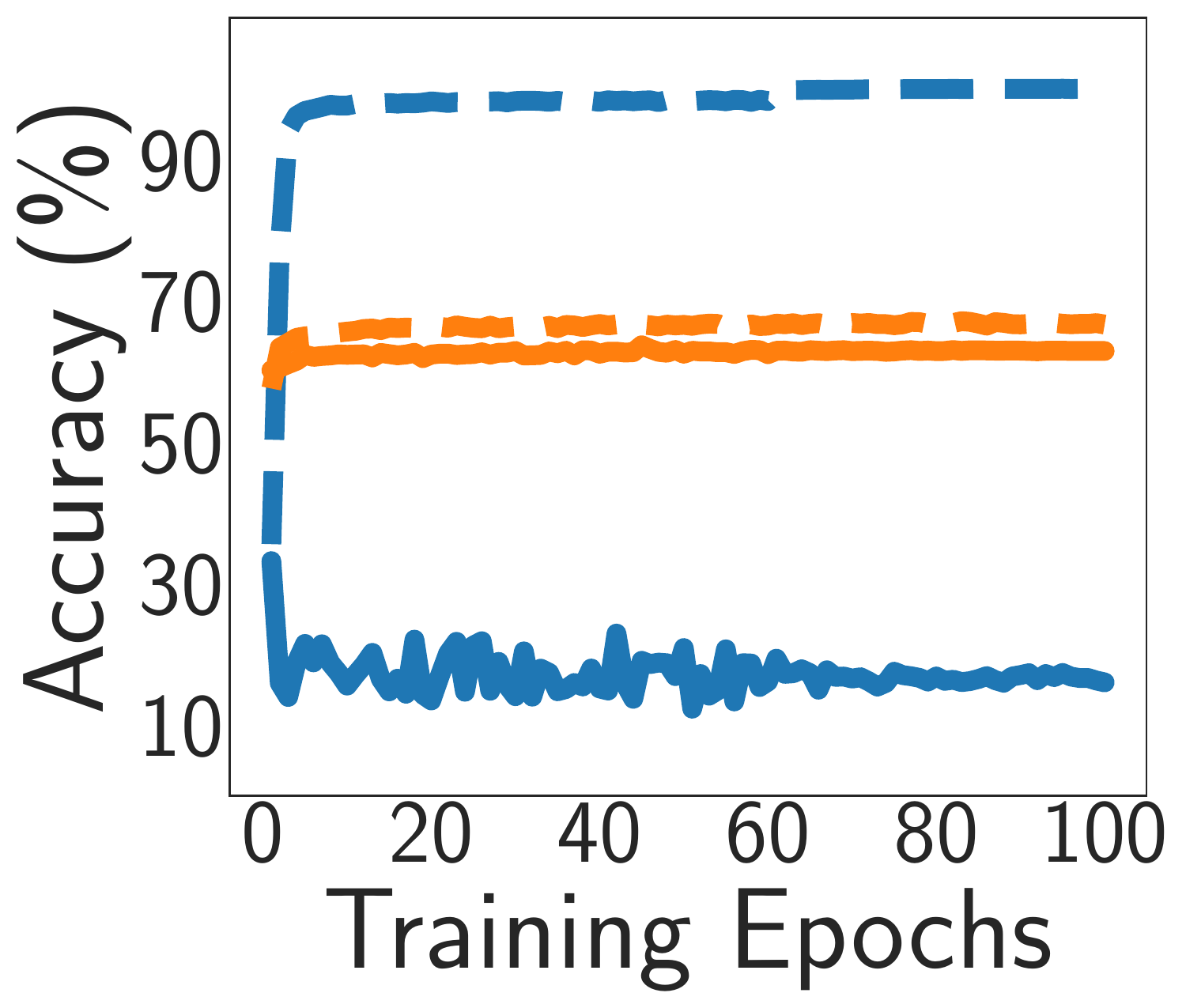}
\caption{HF}
\label{figure:different_training_methods_hf}
\end{subfigure}
\hspace{2mm}
\begin{subfigure}{0.14\columnwidth}
\includegraphics[width=\columnwidth]{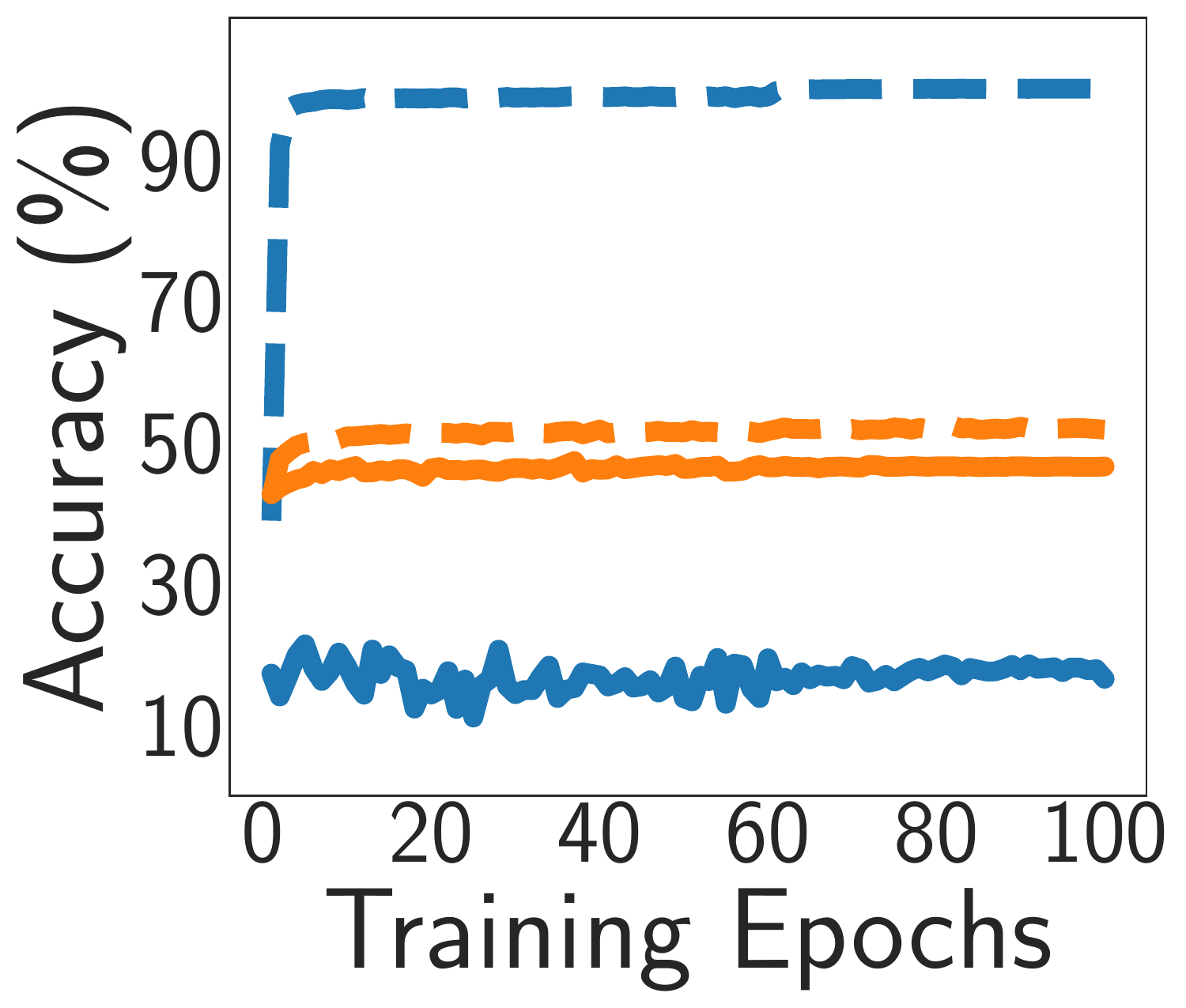}
\caption{CC}
\label{figure:different_training_methods_cc}
\end{subfigure}
\hspace{2mm}
\begin{subfigure}{0.14\columnwidth}
\includegraphics[width=\columnwidth]{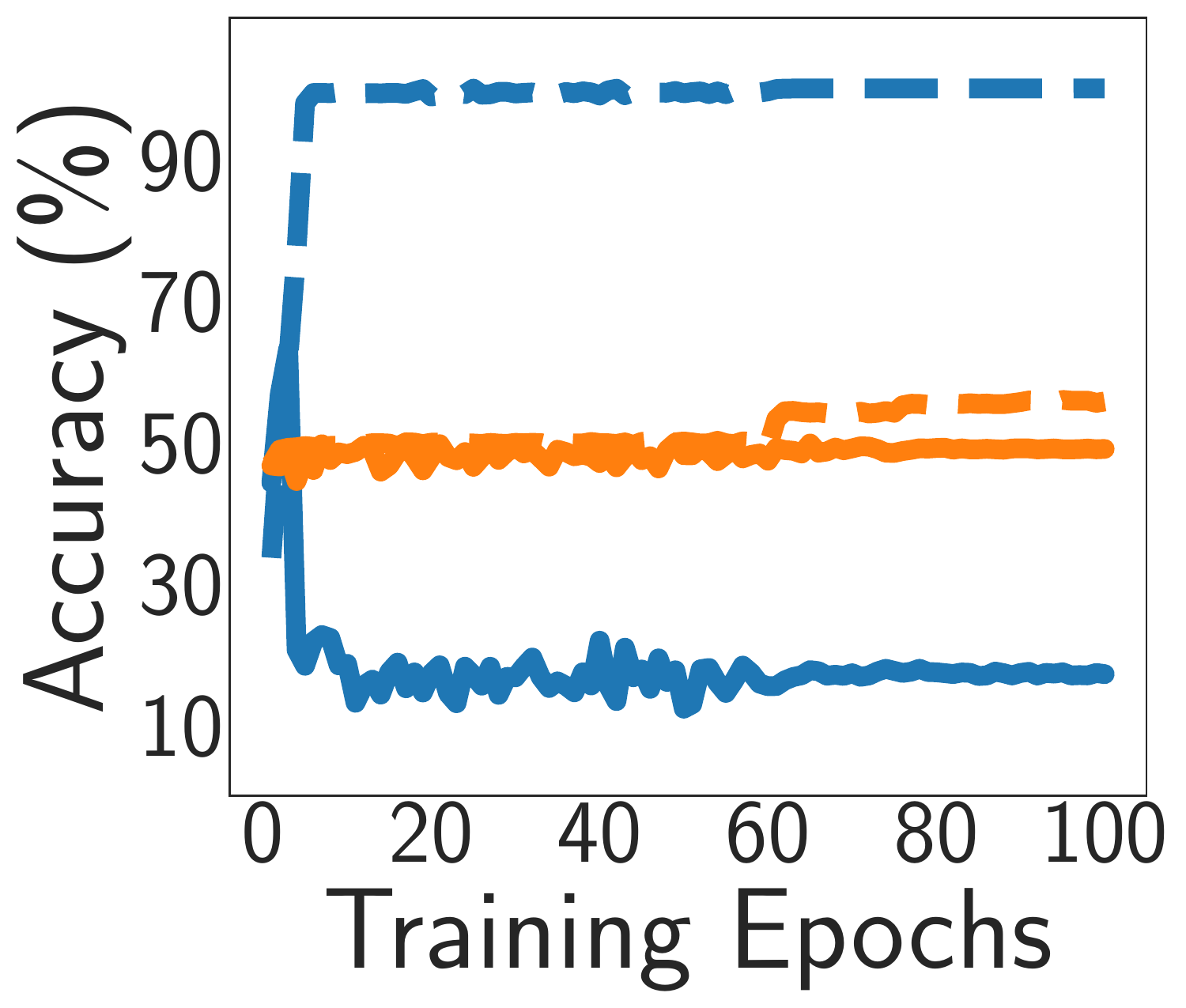}
\caption{Two Models}
\label{figure:different_training_methods_two_models}
\end{subfigure}
\hspace{2mm}
\begin{subfigure}{0.14\columnwidth}
\includegraphics[width=\columnwidth]{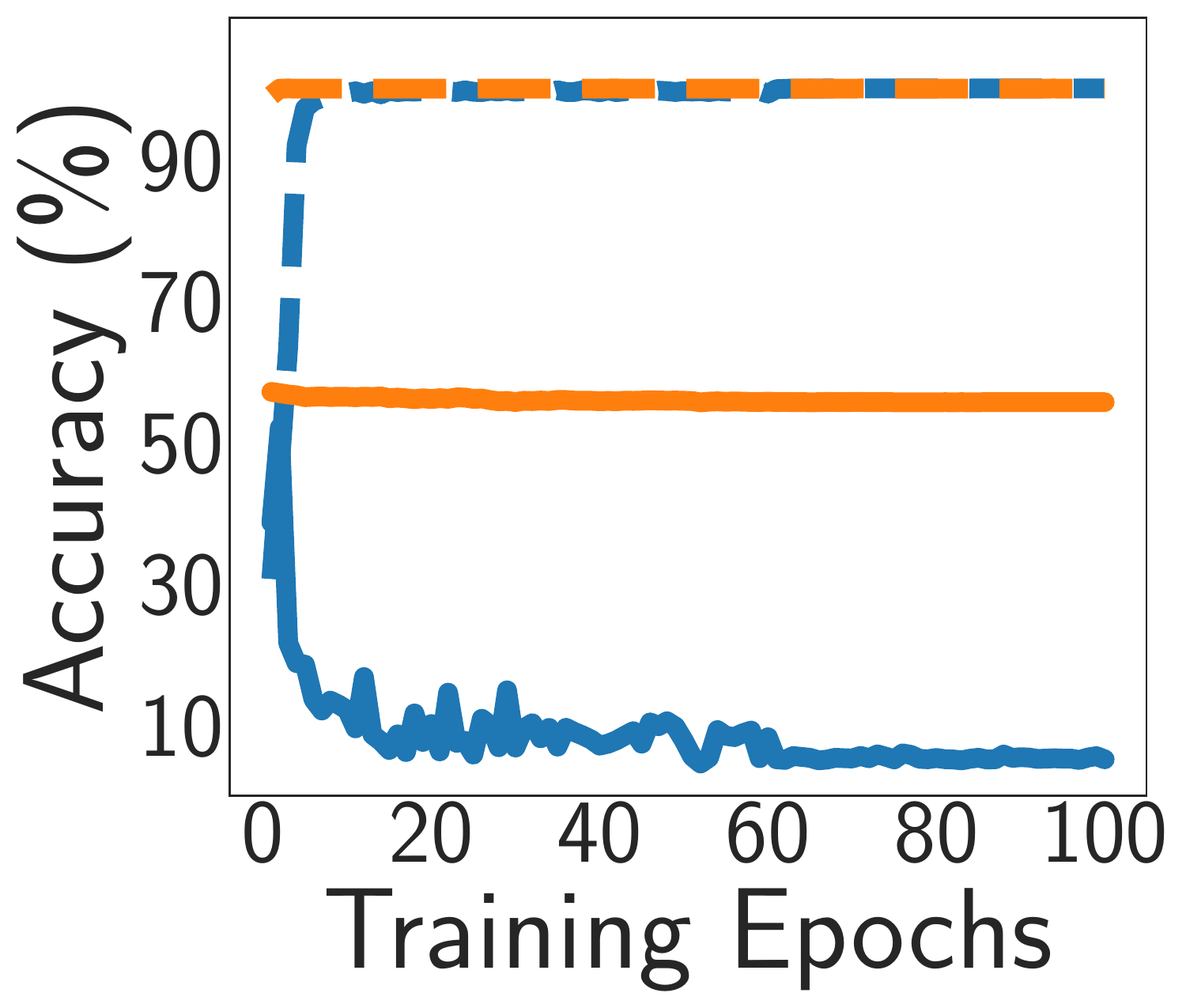}
\caption{TAP-based}
\label{figure:different_training_methods_cptap}
\end{subfigure}
\hspace{2mm}
\begin{subfigure}{0.14\columnwidth}
\includegraphics[width=\columnwidth]{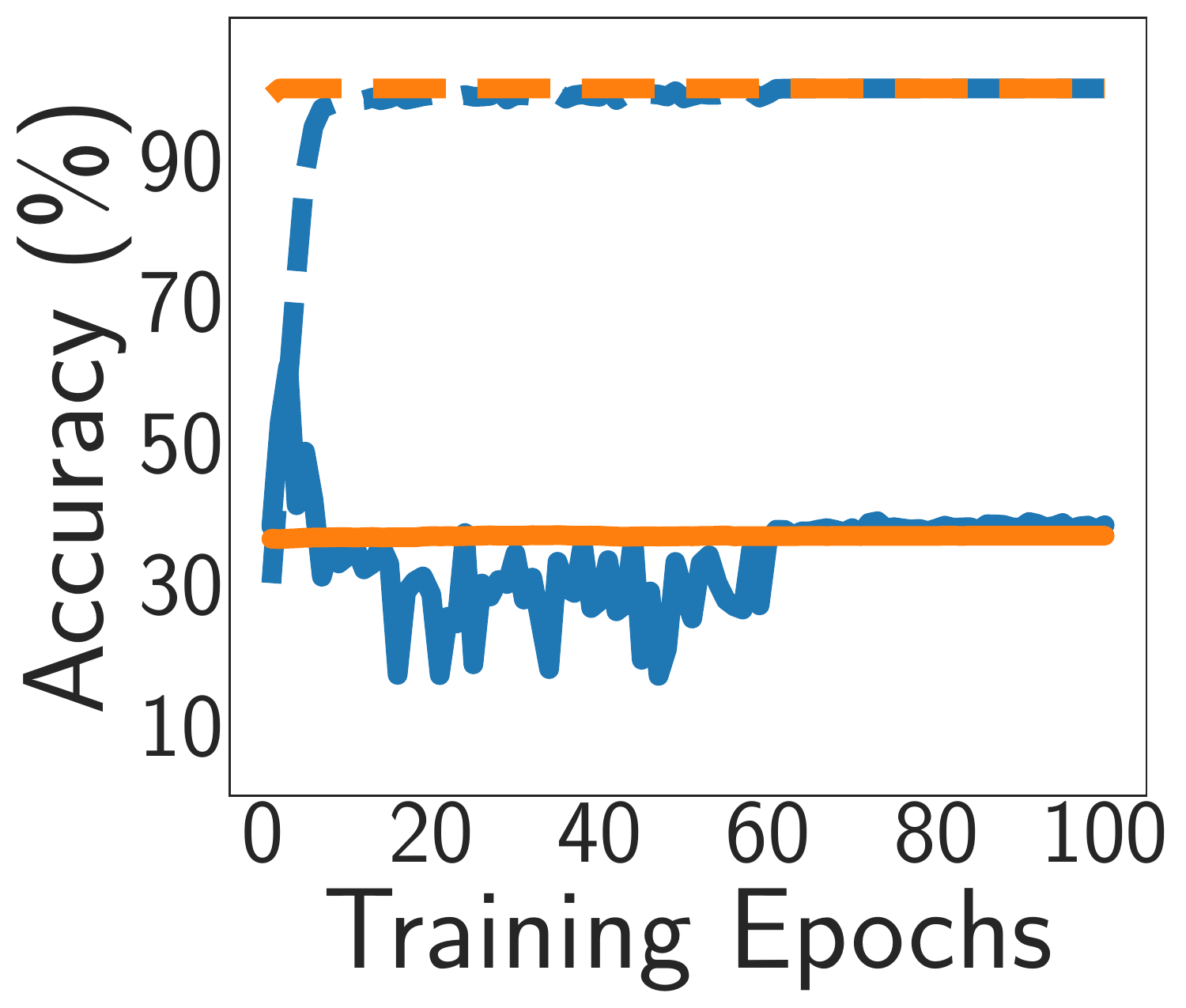}
\caption{Ours}
\label{figure:different_training_methods_ours}
\end{subfigure}
\caption{Training process for supervised learning and linear probing process after the pretraining under SimCLR for clean and various attacking methods on CIFAR-10.}
\label{figure:different_training_methods}
\end{figure*}

In order to understand the working mechanisms of different alternative attacks, we plot the variation of accuracy through the training process for SL and linear probing process after pretraining for SimCLR on CIFAR-10. Figure~\ref{figure:different_training_methods} illustrates that HF, CC and the attack with two models work similarly, as only after a few epochs, SL overfits to the poisoned data, while SimCLR almost keeps the close training and test accuracy through the whole linear probing process. TAP-based attack exhibits similar learning pattern for SL, while the training accuracy of SimCLR is kept close to 100\% along the process. The training characteristics of our \textit{Transferable Poisoning} is more close to TAP-based attack, whereas we find a better way for poisoning both supervised and contrastive learning in terms of the worst-case unlearnability. For HF, CC and the attack with two models, since the poisoning perturbations for supervised and contrastive learning are generated on two unrelated models, one of the perturbations behaves like random noise to the other. It seems the poisons for SL are hardly affected by the poisons from SimCLR, while conversely, the influence is more severe. And as the unlearnability of error-minimization (EM) for supervised learning will be highly blocked by contrastive learning, so the training and test accuracy almost have no changes along the linear probing. For TAP-based attack, though the poisoning effect on SL still exists after contrastive learning as the model quickly overfits to the poisoned data in linear probing, this error-maximization noise contradicts to the poison created for contrastive learning, which is essentially an error-minimization noise, and thus downgrades the unlearnability for contrastive learning. As a comparison, our \textit{Transferable Poisoning} first generates error-minimization noises for both SL and contrastive learning, which has less contradictions to each other. And we also use one shared model for the two training paradigms, which automatically searches the optimization direction for poisoning both of them, thus the final generated poisons have more correlations to each other, and enable much better worst-case unlearnability.

\section{Conclusion}
We study availability poisoning attacks under a more realistic scenario, where the victim has the freedom to choose any possible learning algorithm to reach the goal of their targeted task. We show that existing attacks targeting for a specific learning paradigm exhibit poor transferability in this setting, especially for contrastive learning. In order to make the poisons transfer across different victim learners, we propose to first use alignment and uniformity loss to poison contrastive learning, and then iteratively leverage the gradient information from supervised and contrastive learning to enhance the transferability of the generated poisons. Extensive experiments consistently demonstrate that our method exhibits superior transferability across diverse learning algorithms compared to prior attacks. Our research uncovers that availability poisoning attacks are more severe than previously anticipated, and we hope it can inspire the community to delve deeper into this field in the future.

%-------------------------------------------------------------------------------
\bibliographystyle{plain}
\bibliography{my}

\appendix
\section{Detailed Experimental Settings}
\label{sec:detailed_settings}

\subsection{Datasets Description}
In our experiments, we adopt four commonly used benchmark datasets for classification tasks, i.e., CIFAR-10, CIFAR-100, TinyImageNet and MiniImageNet. Among them, CIFAR-10 has 60000 images for 10 classes and each class has the same number of samples; CIFAR-100 is also a balanced datasets which has 100 classes with 600 number of images for each class, both CIFAR-10 and CIFAR-100 has the size of $32\times32\times3$ for each image. TinyImageNet has 200 classes, and each class has 500 training samples and 50 testing samples, the size for each data sample is $64\times64\times3$. MiniImageNet has 100 classes and 60000 images in total while the size of each image is not fixed, and there isn't exact part for training and testing. So we randomly split 48000 data samples for training and the rest for testing to form a balanced dataset, and crop each image to the same size as $64\times64\times3$.

\subsection{Poison Generation}
The iterative training epochs $T = 300$ and the number of updates in each epoch $M = |\mathcal{D}_{c}|/B$ where $|\mathcal{D}_{c}|$ is the size of the clean dataset and $B$ is the batch size, which means we pass all the training data samples in each epoch. The model $\theta_b + \theta_s$ is updated with SGD optimizer with momentum $0.9$ and weight decay $5\times10^{-4}$ and learning rate $\eta_\mathrm{sl} = 0.1$ for supervised learning, and the model $\theta_b + \theta_c$ is updated with SGD optimizer with momentum $0.9$ and weight decay $1\times10^{-4}$ and learning rate $\eta_\mathrm{cl} = 0.5$ for contrastive learning. $\lambda = 1$, $\alpha = 2$ and $t = 2$ for alignment and uniformity loss. The poisoning perturbations $\bm{\delta}$ is updated based on model $\theta_b + \theta_s$ with number of PGD steps $S_{\mathrm{sl}} = 5$ and step size $\alpha_{\mathrm{sl}} = 0.1\times\epsilon$. And as for the update based on model $\theta_b + \theta_c$, the number of PGD steps is $5$, and the step size $\alpha_{\mathrm{cl}} = 0.1\times\epsilon$ as well. The poison budget $\epsilon = 8/255$.

The cross-entropy loss for supervised learning for each example $(\bm{x}, y)$ is defined as follows:
\begin{equation*}
    \label{equ:cross_entropy}
    \mathcal{L}_{\mathrm{CE}}(f(\bm{x};\theta),y)=-\sum_{i=1}^C y_i \log{[f(\bm{x};\theta)]_{i}},
\end{equation*}
where $C$ is the total number of classes. 
$y_i$ equals $1$ only if the sample belongs to class $i$ and otherwise $0$, and $[f(\bm{x};\theta)]_i$ is the $i$-th element of the prediction posteriors. 

\subsection{Evaluation}
For supervised learning, we train the model for $100$ epochs with SGD optimizer with momentum $0.9$ and weight decay $5\times10^{-4}$, and initial learning rate of $0.1$ which is decayed by a factor of $0.1$ at 60-th, 75-th and 90-th epoch. For supervised contrastive learning, SimCLR, BYOL, MoCov2 and SimSiam, we use the same setting for CIFAR-10 and CIFAR-100, i.e., we first pretrain the model for $1000$ epochs, the otimizer is SGD with momentum $0.9$ and weight decay $1\times10^{-4}$, the initial learning rate is $0.5$ for supervised contrastive learning, SimCLR and SimSiam, $1.0$ for BYOL and $0.3$ for MoCov2, and they are decayed by a cosine scheduler, the temperature for the loss in supervised contrastive learning and SimCLR is $0.5$, and $0.2$ for MoCov2. We then train the linear classifier for $100$ epochs with SGD optimizer with momentum $0.9$ and weight decay $0$, and initial learning rate of $1.0$ which is decayed by a factor of $0.2$ at 60-th, 75-th and 90-th epoch. For TinyImageNet and MiniImageNet, all the settings are the same as for CIFAR-10 and CIFAR-100, except for that we change the initial learning rate in pretraining to $0.15$ for supervised contrastive learning and SimCLR, $0.3$ for BYOL and $0.1$ for SimSiam.

\section{Additional Results}
\label{sec:additional_results}

\subsection{Defense}
We evaluate the attack performance against different defenses, mainly including adversarial training: AT~\cite{MMSTV18} for supervised learning and AdvCL~\cite{KTH20} for contrastive learning, and data augmentations: Ueraser~\cite{QGZYX23} and ISS~\cite{LZL23} for supervised learning, and Cutout~\cite{DT17} and Gaussian Blur for contrastive learning. As demonstrated in Table~\ref{table:defense}, for supervised learning, ISS-JPEG and AT are more effective as they can recover the accuracy to a similar level for all the attacks, while they will also downgrade the accuracy with clean data to some extent. For contrastive learning, Gaussian Blur is the best one among the adopted defenses as it not only slightly improves the accuracy with clean data, but also mitigates the unlearnability of all the attacks. Same as the claims in Section~\ref{sec:experiment}, our attack still shows more advantage on contrastive learning, and aligns the attack performance on both supervised and contrastive learning.

\subsection{Computational Cost}
We show the training time of each attack for generating the poisoning perturbations using a single NVIDIA DGX A100 in Table~\ref{table:computational_cost}. Our attack can be more efficient than CP as we use less training epochs. While as our attack also needs to poison contrastive learning, it costs more computational resources than other attacks, and we view it as future work to make our attack more efficient.

\subsection{Poison Budget}
We apply different poison budget for generating the perturbations and evaluate the attack performance. Table~\ref{table:other_poisoning_budgets} demonstrates that the test accuracy is a function of the poison budget for all the adopted training algorithms, as the accuracy decreases with the poison budget increases.

\subsection{Visualization}
\begin{figure*}[!t]
\centering
\hspace{-4mm}
\subfloat[EM]{\includegraphics[width=0.15\columnwidth]{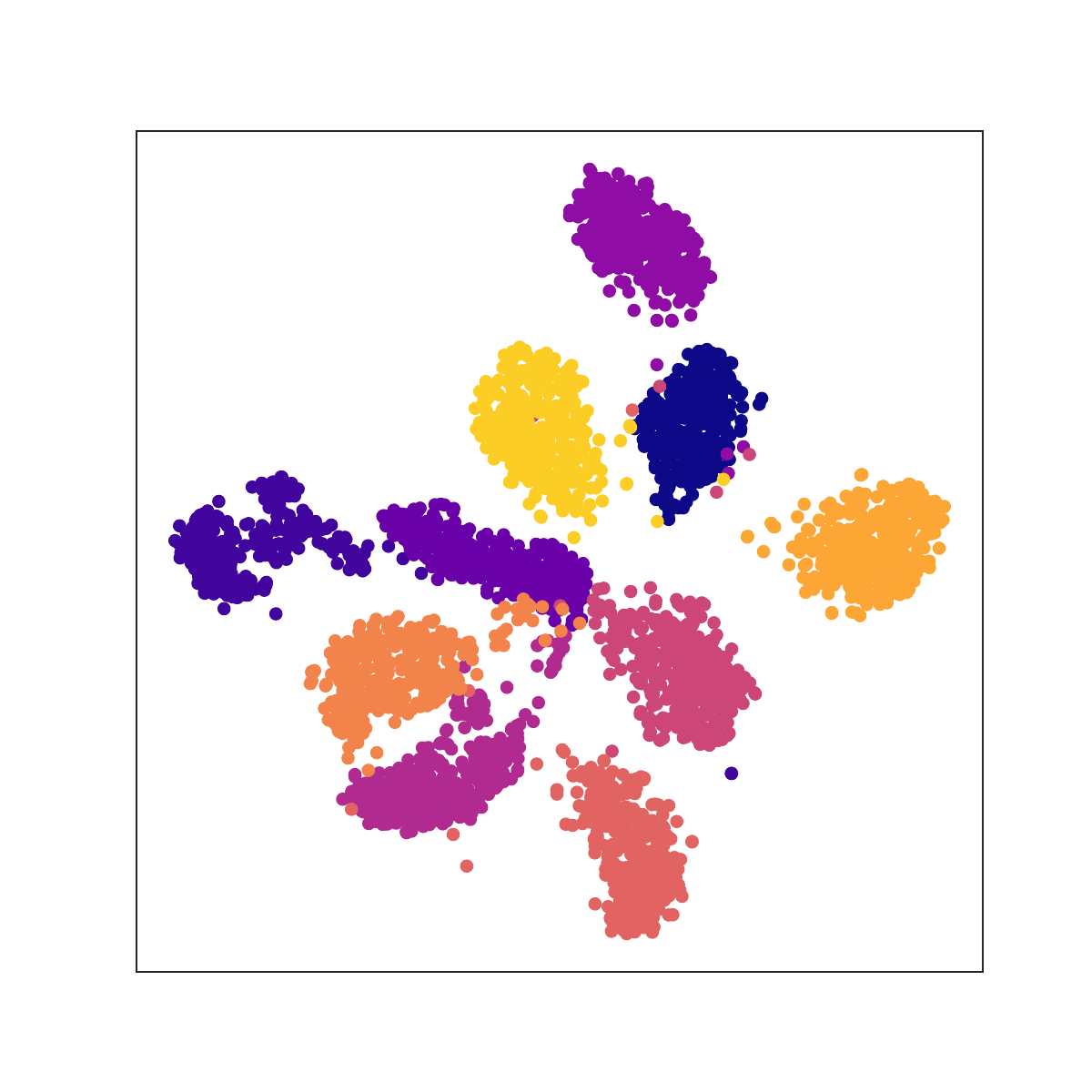}
\label{figure:visualization_em}}
\hspace{1mm}
\subfloat[TAP]{\includegraphics[width=0.15\columnwidth]{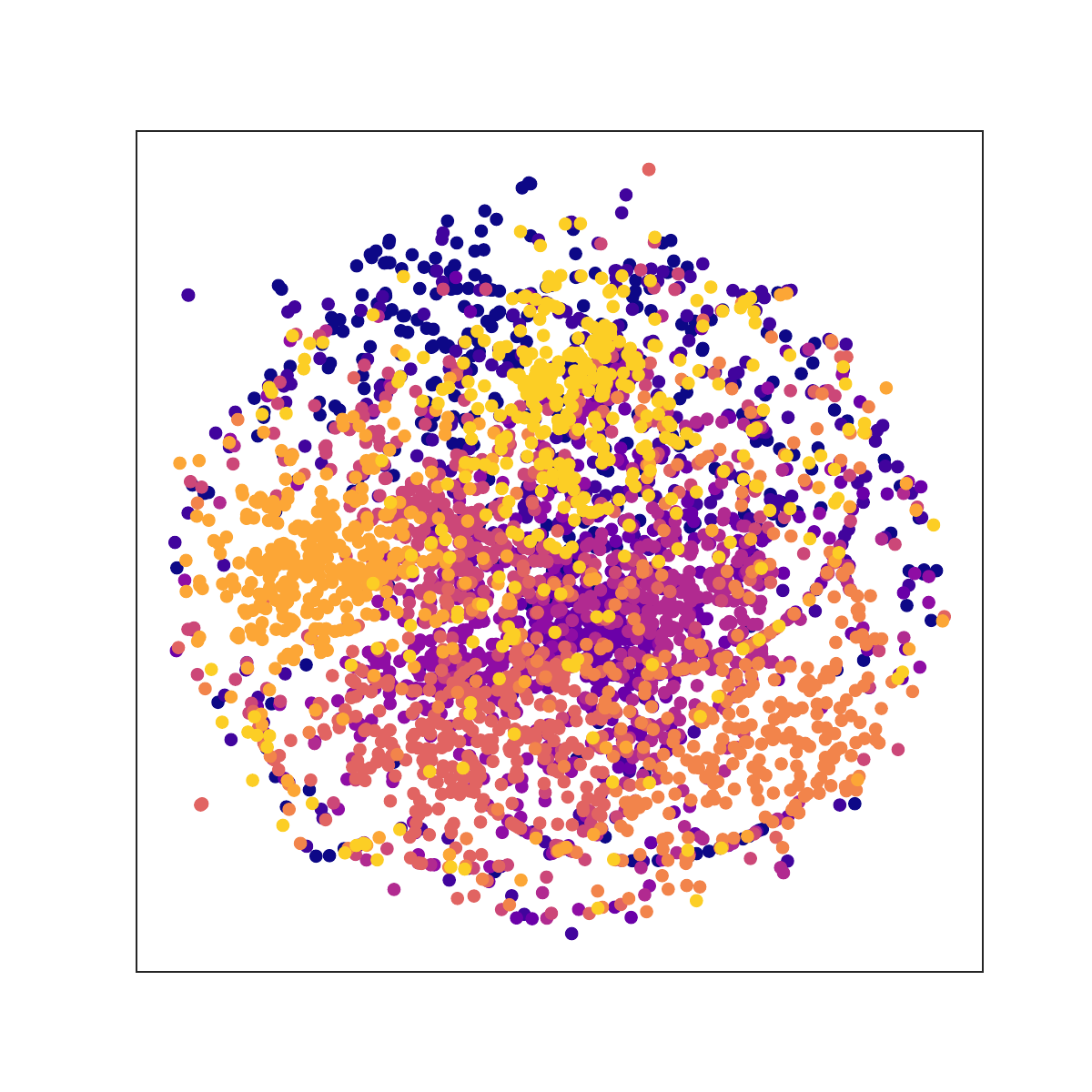}
\label{figure:visualization_tap}}
\hspace{1mm}
\subfloat[CP-BYOL]{\includegraphics[width=0.15\columnwidth]{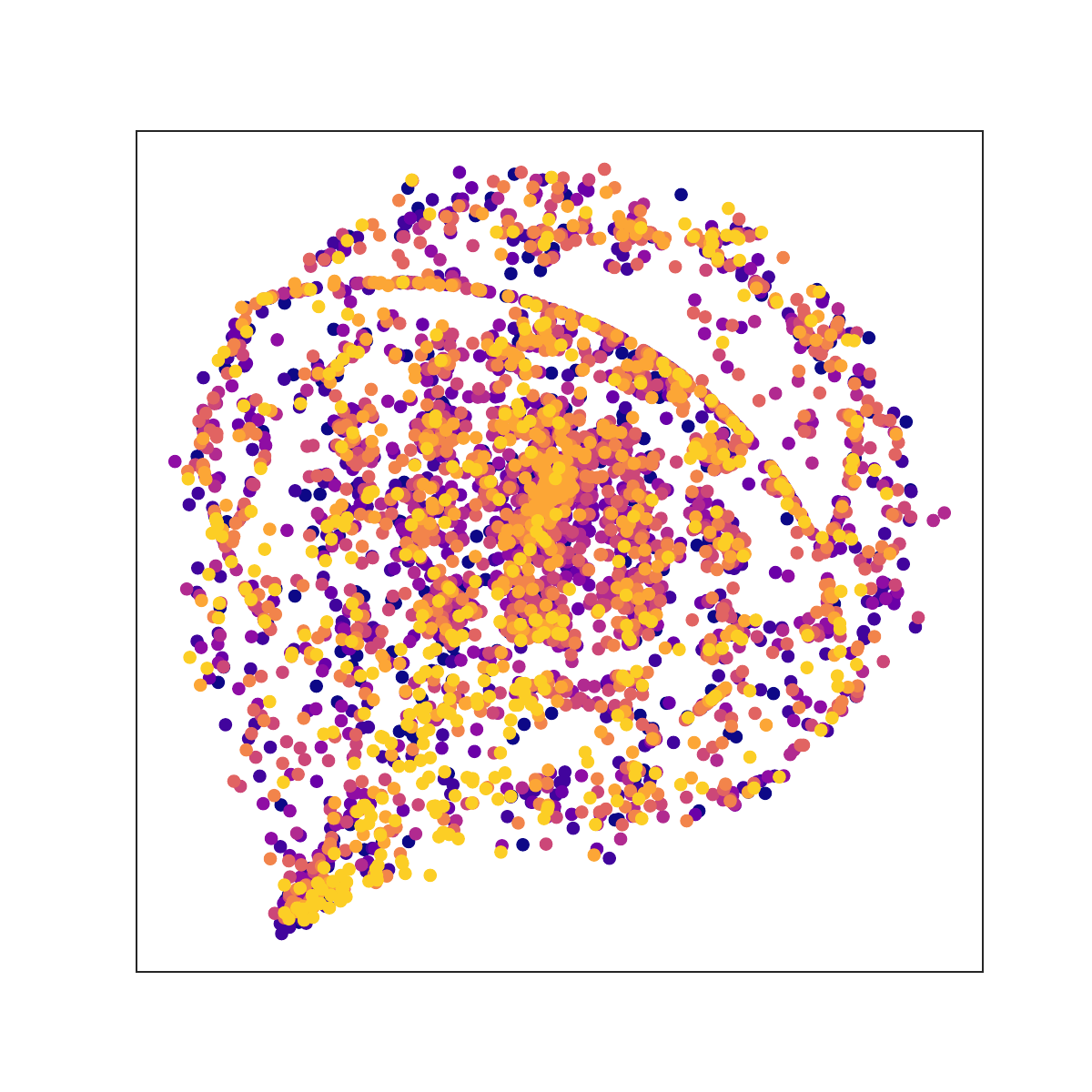}
\label{figure:visualization_cp_byol}}
\hspace{1mm}
\subfloat[TUE]{\includegraphics[width=0.15\columnwidth]{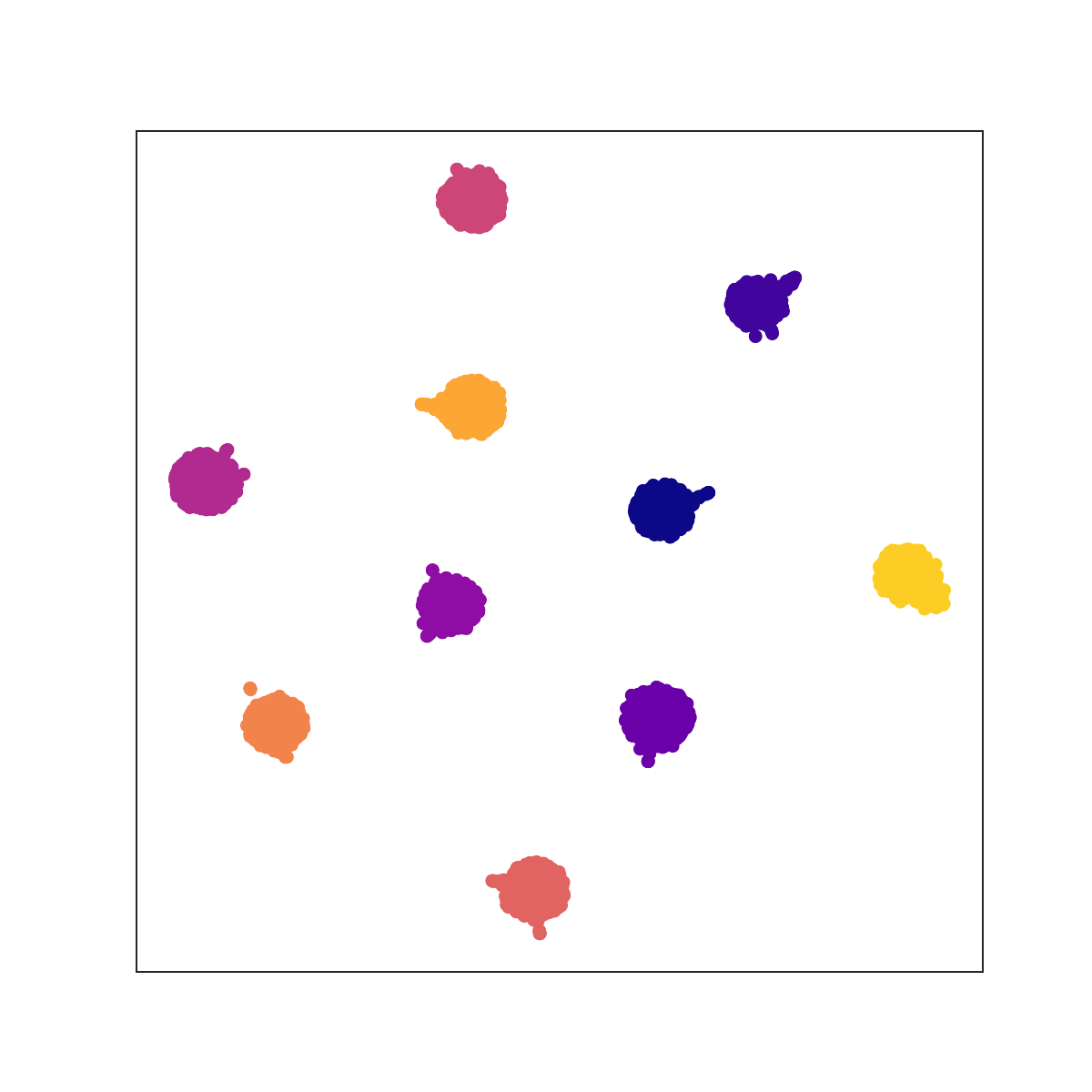}
\label{figure:visualization_tue}}
\hspace{1mm}
\subfloat[T-AAP]{\includegraphics[width=0.15\columnwidth]{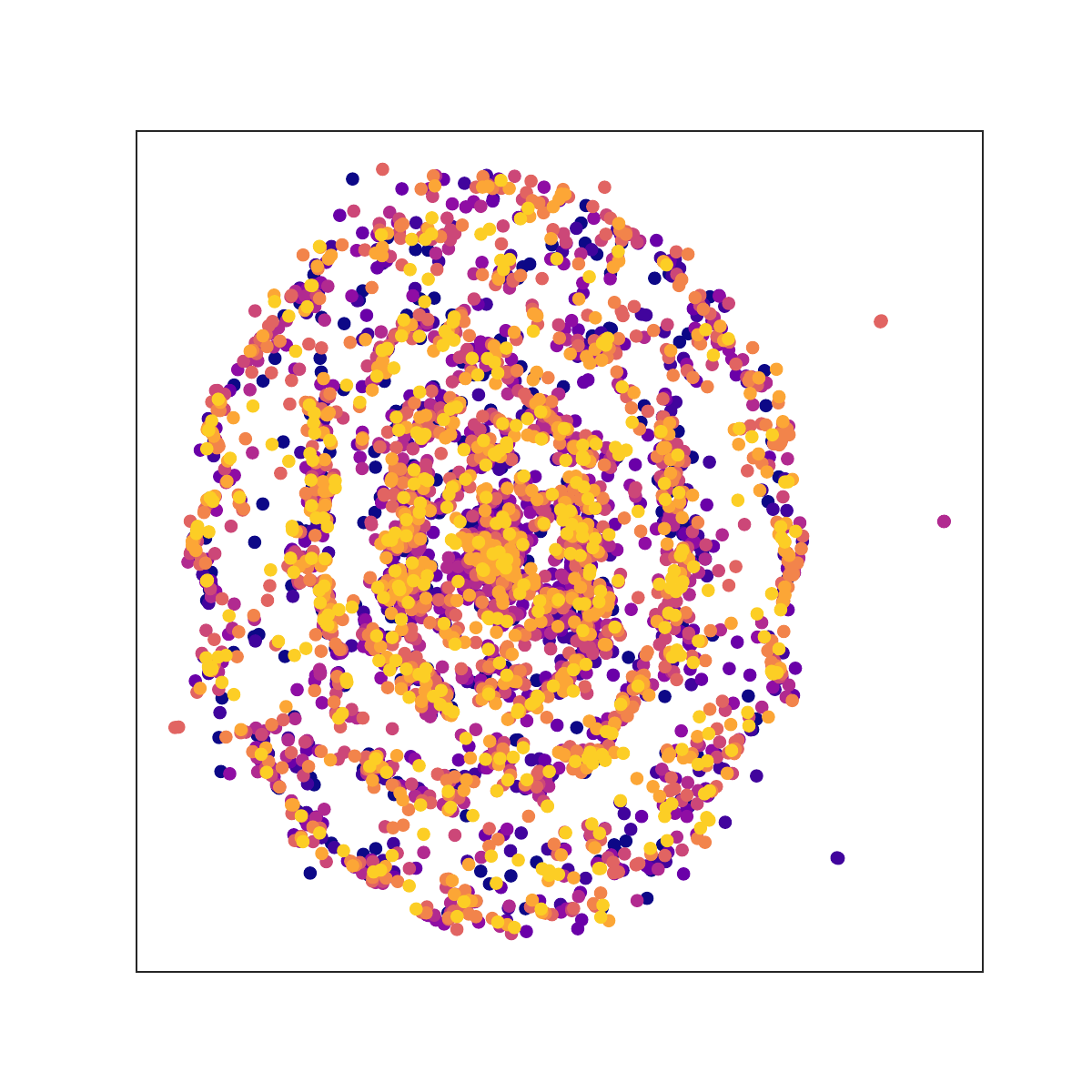}
\label{figure:visualization_aap}}
\hspace{1mm}
\subfloat[Ours]{\includegraphics[width=0.15\columnwidth]{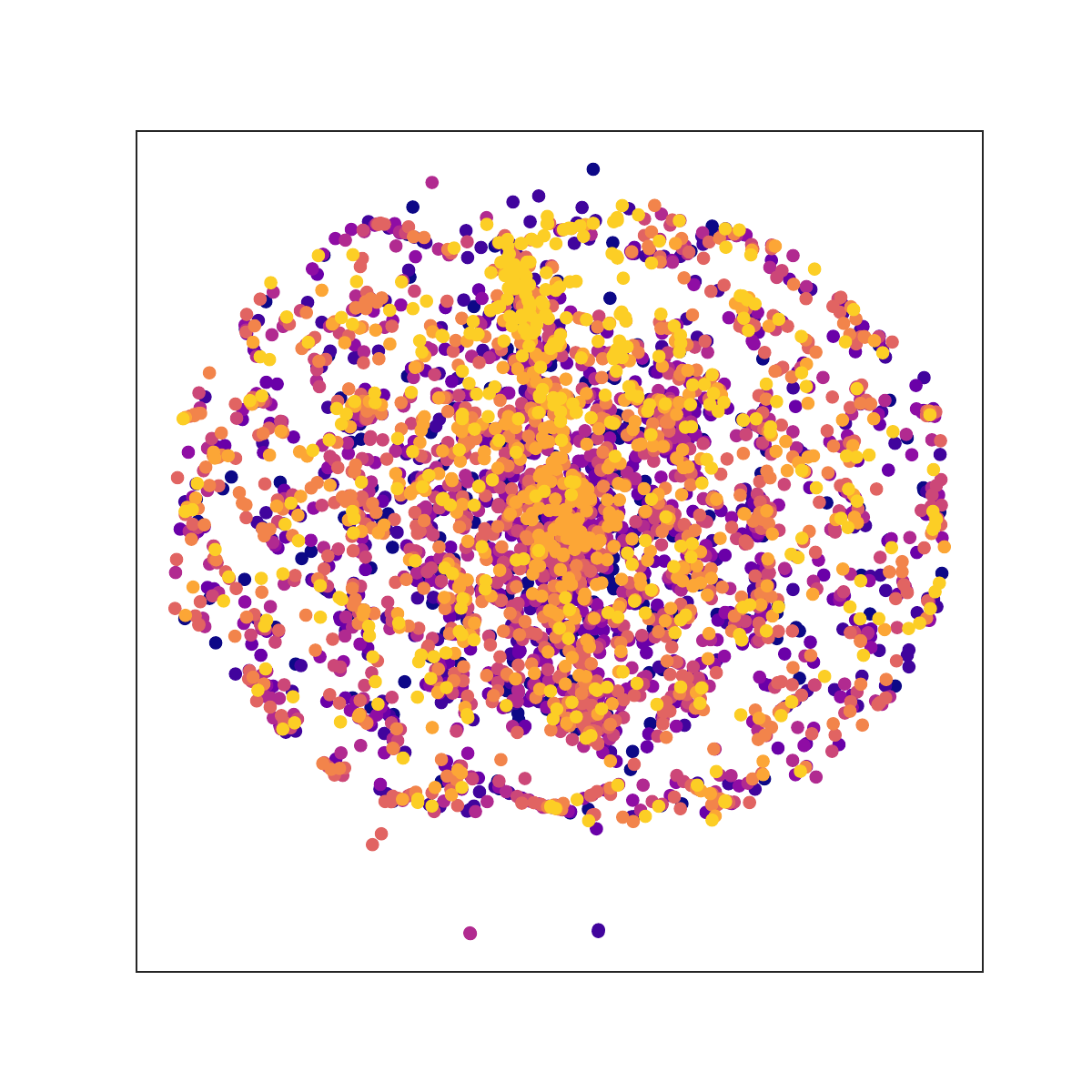}
\label{figure:visualization_ours}}
\caption{t-SNE of the generated generated noises from EM, TAP, CP-BYOL, TUE, T-AAP and Ours on CIFAR-10.}
\vspace{-2mm}
\label{figure:visualization_baseline}
\end{figure*}

\begin{figure}[h]
\centering
\captionsetup{type=figure}
\begin{subfigure}{1.0\columnwidth}
\includegraphics[width=0.09\columnwidth]{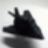}
\includegraphics[width=0.09\columnwidth]{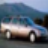}
\includegraphics[width=0.09\columnwidth]{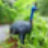}
\includegraphics[width=0.09\columnwidth]{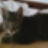}
\includegraphics[width=0.09\columnwidth]{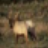}
\includegraphics[width=0.09\columnwidth]{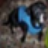}
\includegraphics[width=0.09\columnwidth]{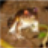}
\includegraphics[width=0.09\columnwidth]{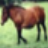}
\includegraphics[width=0.09\columnwidth]{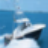}
\includegraphics[width=0.09\columnwidth]{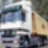}
\end{subfigure}
\vspace{1mm}
\begin{subfigure}{1.0\columnwidth}
\includegraphics[width=0.09\columnwidth]{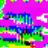}
\includegraphics[width=0.09\columnwidth]{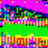}
\includegraphics[width=0.09\columnwidth]{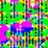}
\includegraphics[width=0.09\columnwidth]{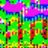}
\includegraphics[width=0.09\columnwidth]{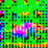}
\includegraphics[width=0.09\columnwidth]{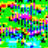}
\includegraphics[width=0.09\columnwidth]{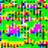}
\includegraphics[width=0.09\columnwidth]{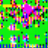}
\includegraphics[width=0.09\columnwidth]{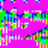}
\includegraphics[width=0.09\columnwidth]{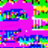}
\end{subfigure}
\vspace{1mm}
\begin{subfigure}{1.0\columnwidth}
\includegraphics[width=0.09\columnwidth]{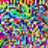}
\includegraphics[width=0.09\columnwidth]{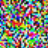}
\includegraphics[width=0.09\columnwidth]{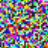}
\includegraphics[width=0.09\columnwidth]{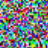}
\includegraphics[width=0.09\columnwidth]{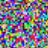}
\includegraphics[width=0.09\columnwidth]{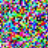}
\includegraphics[width=0.09\columnwidth]{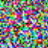}
\includegraphics[width=0.09\columnwidth]{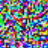}
\includegraphics[width=0.09\columnwidth]{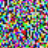}
\includegraphics[width=0.09\columnwidth]{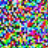}
\end{subfigure}
\vspace{1mm}
\begin{subfigure}{1.0\columnwidth}
\includegraphics[width=0.09\columnwidth]{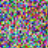}
\includegraphics[width=0.09\columnwidth]{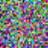}
\includegraphics[width=0.09\columnwidth]{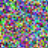}
\includegraphics[width=0.09\columnwidth]{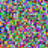}
\includegraphics[width=0.09\columnwidth]{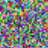}
\includegraphics[width=0.09\columnwidth]{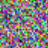}
\includegraphics[width=0.09\columnwidth]{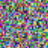}
\includegraphics[width=0.09\columnwidth]{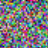}
\includegraphics[width=0.09\columnwidth]{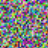}
\includegraphics[width=0.09\columnwidth]{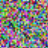}
\end{subfigure}
\vspace{1mm}
\begin{subfigure}{1.0\columnwidth}
\includegraphics[width=0.09\columnwidth]{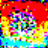}
\includegraphics[width=0.09\columnwidth]{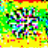}
\includegraphics[width=0.09\columnwidth]{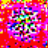}
\includegraphics[width=0.09\columnwidth]{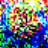}
\includegraphics[width=0.09\columnwidth]{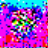}
\includegraphics[width=0.09\columnwidth]{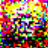}
\includegraphics[width=0.09\columnwidth]{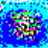}
\includegraphics[width=0.09\columnwidth]{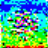}
\includegraphics[width=0.09\columnwidth]{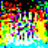}
\includegraphics[width=0.09\columnwidth]{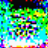}
\end{subfigure}
\vspace{1mm}
\begin{subfigure}{1.0\columnwidth}
\includegraphics[width=0.09\columnwidth]{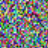}
\includegraphics[width=0.09\columnwidth]{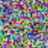}
\includegraphics[width=0.09\columnwidth]{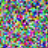}
\includegraphics[width=0.09\columnwidth]{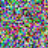}
\includegraphics[width=0.09\columnwidth]{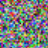}
\includegraphics[width=0.09\columnwidth]{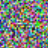}
\includegraphics[width=0.09\columnwidth]{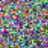}
\includegraphics[width=0.09\columnwidth]{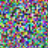}
\includegraphics[width=0.09\columnwidth]{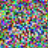}
\includegraphics[width=0.09\columnwidth]{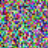}
\end{subfigure}
\vspace{1mm}
\hspace{-7mm}
\begin{subfigure}{0.09\columnwidth}
\includegraphics[width=1.0\columnwidth]{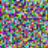}
\caption{plane}
\end{subfigure}
\begin{subfigure}{0.09\columnwidth}
\includegraphics[width=1.0\columnwidth]{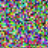}
\caption{car}
\end{subfigure}
\begin{subfigure}{0.09\columnwidth}
\includegraphics[width=1.0\columnwidth]{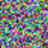}
\caption{bird}
\end{subfigure}
\begin{subfigure}{0.09\columnwidth}
\includegraphics[width=1.0\columnwidth]{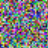}
\caption{cat}
\end{subfigure}
\begin{subfigure}{0.09\columnwidth}
\includegraphics[width=1.0\columnwidth]{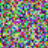}
\caption{deer}
\end{subfigure}
\begin{subfigure}{0.09\columnwidth}
\includegraphics[width=1.0\columnwidth]{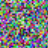}
\caption{dog}
\end{subfigure}
\begin{subfigure}{0.09\columnwidth}
\includegraphics[width=1.0\columnwidth]{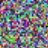}
\caption{frog}
\end{subfigure}
\begin{subfigure}{0.09\columnwidth}
\includegraphics[width=1.0\columnwidth]{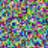}
\caption{horse}
\end{subfigure}
\begin{subfigure}{0.09\columnwidth}
\includegraphics[width=1.0\columnwidth]{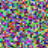}
\caption{ship}
\end{subfigure}
\begin{subfigure}{0.09\columnwidth}
\includegraphics[width=1.0\columnwidth]{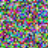}
\caption{truck}
\end{subfigure}
\caption{Visualizations of perturbations for CIFAR-10 images with one for each class. From top to bottom are the original images, noises generated by EM, TAP, CP-BYOL, TUE, T-AAP and Ours.}
\label{fig:additional visulization noises}
\end{figure}

We show the t-SNE and perturbations for each class on CIFAR-10 dataset for EM, TAP, CP-BYOL, TUE, T-AAP and our attack, and the results are in Figure~\ref{figure:visualization_baseline} and~\ref{fig:additional visulization noises} respectively. EM and TUE exhibit obvious linear separation and more simple perturbation patterns, which makes them quite ineffective for poisoning contrastive learning. For the rest attack, our method seems contain more complex perturbation patterns, especially for the result in t-SNE.

\begin{table*}[t]
\centering
\caption{Attack performance against defenses on CIFAR-10. SimCLR is adopted for CL.}
\vspace{-2mm}
\setlength{\tabcolsep}{6.0pt}
\label{table:defense}
\begin{tabular}{ll|ccccc|c}
\toprule
\rowcolor{white}
& Defense & Clean & EM & TAP & TUE & T-AAP & Ours\\
\midrule
 & No Defense & 94.59 & 18.47 & 6.94 & 29.37 & 10.44 & 37.34 \\
 & UEraser-max & 79.88 & 72.02 & 61.80 & 67.98 & 63.25 & 73.27 \\
SL & ISS-Grayscale & 92.97 & 91.09 & 9.97 & 27.31 & 9.87 & 40.61 \\
 & ISS-JPEG & 85.15 & 81.30 & 84.65 & 82.56 & 84.71 & 82.43 \\
 & AT & 85.06 & 84.67 & 83.22 & 84.08 & 83.04 & 84.42 \\ 
\midrule
 & No Defense & 91.18 & 89.31 & 76.66 & 88.31 & 48.61 & 38.61 \\
 & Random Noise & 90.49 & 89.97 & 61.96 & 84.29 & 59.11 & 40.34 \\
CL & Cutout & 92.56 & 89.59 & 57.45 & 82.09 & 52.90 & 33.34 \\
 & Gaussian Blur & 91.25 & 88.23 & 78.08 & 86.99 & 81.10 & 80.03 \\
 & AdvCL & 81.22 & 80.85 & 79.28 & 79.47 & 74.83 & 72.41 \\
\bottomrule
\end{tabular}
\end{table*}

\begin{table*}[t]
\centering
\caption{The time (h) for generating the poisoning perturbations of CIFAR-10.}
\vspace{-2mm}
\setlength{\tabcolsep}{8.0pt}
\label{table:computational_cost}
\begin{tabular}{l|cccccc}
\toprule
\rowcolor{white}
Attack & EM & TAP & CP-BYOL & TUE & T-AAP & Ours\\
\midrule
 Time & 0.21 & 0.63 & 39.42 & 2.15 & 1.52 & 15.61 \\ 
\bottomrule
\end{tabular}
\end{table*}

\begin{table*}[t]
\centering
\caption{Attack performance of our method under different poison budget.}
\vspace{-2mm}
\setlength{\tabcolsep}{4.0pt}
\label{table:other_poisoning_budgets}
\begin{tabular}{l|c|ccccc|c}
\toprule
\rowcolor{white}
poisoning rate & SL & SupCL & SimCLR & BYOL & MoCo & SimSiam & Worst\\
\midrule
0 & 94.59 & 94.75 & 91.18 & 92.00 & 91.43 & 90.59 & 94.75  \\
4/255 & 50.77 & 61.65 & 54.54 & 51.63 & 46.97 & 30.40 & 61.65 \\
8/255 & 37.34 & 41.72 & 35.43 & 38.61 & 30.98 & 15.58 & 41.72 \\
16/255 & 32.73 & 32.90 & 23.26 & 24.20 & 20.74 & 12.55 & 32.90 \\
\bottomrule
\end{tabular}
\end{table*}

\end{document}